\documentclass{emulateapj}
\usepackage{graphicx}

\shorttitle{SN 2011dh: UV to NIR}
\shortauthors{Marion et al.}

\newcommand{\h}{\ion{H}{1}}
\newcommand{\ha}{H$\alpha$}
\newcommand{\he}{\ion{He}{1}}
\newcommand{\ca}{\ion{Ca}{2}}
\newcommand{\mg}{\ion{Mg}{2}}
\newcommand{\si}{\ion{Si}{2}}
\newcommand{\fe}{\ion{Fe}{2}}
\newcommand{\oi}{\ion{O}{1}}
\newcommand{\ti}{\ion{Ti}{2}}
\newcommand{\co}{\ion{Co}{2}}
\newcommand{\kms}{km s$^{-1}$}

\newcommand{\mum}{$\mu$m}
\newcommand{\about}{$\approx$~}

\newcommand{\bmax}{\emph{B}-max}

\begin{document} 

\title{Type II\lowercase{b} Supernova SN 2011\lowercase{dh}:\\
       Spectra and Photometry from the Ultraviolet to the Near-Infrared}

\author{
G.~H.~Marion\altaffilmark{1,2},
Jozsef~Vinko\altaffilmark{2,3},
Robert~P.~Kirshner\altaffilmark{1},
Ryan~J.~Foley\altaffilmark{1} \footnote{Clay Fellow, Harvard-Smithsonian Center for Astrophysics},\\
Perry~Berlind\altaffilmark{1},
Allyson~Bieryla\altaffilmark{1},
Joshua~S.~Bloom\altaffilmark{4},
Michael~L.~Calkins\altaffilmark{1},
Peter~Challis\altaffilmark{1},\\
Roger~A.~Chevalier\altaffilmark{5},
Ryan~Chornock\altaffilmark{1},
Chris~Culliton\altaffilmark{6},
Jason~L.~Curtis\altaffilmark{6},
Gilbert~A.~Esquerdo\altaffilmark{1},
Mark~E.~Everett\altaffilmark{7},
Emilio~E.~Falco\altaffilmark{1},
Kevin~France\altaffilmark{8},
Claes~Fransson\altaffilmark{9},
Andrew~S.~Friedman\altaffilmark{1},
Peter~Garnavich\altaffilmark{10},
Bruno~Leibundgut\altaffilmark{11},
Samuel~Meyer\altaffilmark{1},
Nathan~Smith\altaffilmark{12},
Alicia~M.~Soderberg\altaffilmark{1},
Jesper~Sollerman\altaffilmark{10},
Dan~L.~Starr\altaffilmark{5},
Tamas~Szklenar\altaffilmark{1},
Katalin~Takats\altaffilmark{3,14}
and J.~Craig~Wheeler\altaffilmark{2}}

\altaffiltext{1}{Harvard-Smithsonian Center for Astrophysics, 60 Garden St., Cambridge, MA 02138, USA; \email{gmarion@cfa.harvard.edu}}
\altaffiltext{2}{University of Texas at Austin, 1 University Station C1400, Austin, TX, 78712-0259, USA}
\altaffiltext{3}{Department of Optics and Quantum Electronics, University of Szeged, Domter 9, 6720, Szeged, Hungary}
\altaffiltext{4}{Department of Astronomy, University of California, Berkeley, CA 94720-3411, USA}
\altaffiltext{5}{Astronomy Department, University of Virginia, Charlottesville, VA 22904, USA}
\altaffiltext{6}{Department of Astronomy and Astrophysics, Pennsylvania State University, 525 Davey Laboratory, University Park, PA 16802, USA}
\altaffiltext{7}{National Optical Astronomy Observatory, Tucson, AZ 85719, USA}
\altaffiltext{8}{Center for Astrophysics and Space Astronomy, University of Colorado, 389 UCB, Boulder, CO 80309, USA}
\altaffiltext{9}{Department of Astronomy, The Oskar Klein Centre, Stockholm University, S-106 91 Stockholm, Sweden}
\altaffiltext{10}{Department of Physics, University of Notre Dame, 225 Nieuwland Science Hall, Notre Dame, IN, 46556, USA}
\altaffiltext{11}{European Southern Observatory, 85748 Garching bei M\"unchen, Germany}
\altaffiltext{12}{University of Arizona, Steward Observatory, Tucson, Arizona 85721, USA}
\altaffiltext{13}{Departamento de Ciencias Fisicas, Universidad Andres Bello, Avda. Republica 252, Santiago, Chile}

\begin{abstract}
We report spectroscopic and photometric observations of the Type~IIb SN~2011dh obtained between 4 and 34 days after the estimated date of explosion (May 31.5 UT).   The data cover a wide wavelength range from 2,000~\AA\ in the ultraviolet (UV) to 2.4~\mum\ in the near-infrared (NIR).  Optical spectra provide line profiles and velocity measurements of \h, \he, \ca\ and \fe\ that trace the composition and kinematics of the SN.  NIR spectra show that helium is present in the atmosphere as early as 11 days after the explosion.   A UV spectrum obtained with the Space Telescope Imaging Spectrograph reveals that the UV flux for SN~2011dh is low compared to other SN~IIb.   Modeling the spectrum with SYNOW suggests that the UV deficit is due to line blanketing from \ti\ and \co.  The \h\ and \he\ velocities in SN~2011dh are separated by about 4,000 \kms\ at all phases.  A velocity gap is consistent with models for a pre-explosion structure in which a hydrogen-rich shell surrounds the progenitor.   We estimate that the H-shell of SN~2011dh is \about 8 times less massive than the shell of SN~1993J and \about 3 times more massive than the shell of SN~2008ax.  Light curves (LC) for twelve passbands are presented: \emph{UVW2,UVM2,UVW1,U,u',B,V,r',i',J,H} and \emph{$K_s$}.  In the \emph{B}-band, SN~2011dh reached peak brightness of 13.17~mag at $20.0 \pm 0.5$ days after the explosion.  The maximum bolometric luminosity of $1.8 \pm 0.2 \times 10^{42}$ erg s$^{-1}$ occurred \about 22 days after the explosion.  NIR emission provides more than 30\% of the total bolometric flux at the beginning of our observations and the NIR contribution increases to nearly 50\% of the total by day 34.  The UV produces 16\% of the total flux on day 4, 5\% on day 9 and 1\% on day 34.  We compare the bolometric light curves of SN~2011dh, SN~2008ax and SN~1993J.  The LC are very different for the first twelve days after the explosions but all three SN~IIb display similar peak luminosities, times of peak, decline rates and colors after maximum.  This suggests that the progenitors of these SN~IIb may have had similar compositions and masses but they exploded inside hydrogen shells that that have a wide range of masses.  SN~2011dh was well observed and a likely progenitor star has been identified in pre-explosion images. The detailed observations presented here will help evaluate theoretical models for this supernova and lead to a better understanding of SN~IIb.
\end{abstract}


\section{Introduction}
SN~2011dh (= PTF11eon) was discovered in the nearby ``Whirlpool'' galaxy M~51 ($D \approx 8.05$ Mpc) less than a day after the explosion.  The early detection and close proximity of SN~2011dh provide optimal conditions for detailed observations of the supernova.  

Core-collapse supernovae (SN) are classified according to their observational characteristics.  Type~II~SN are hydrogen rich and their spectra display prominent Balmer series features.  SN~Ib do not have hydrogen features in their spectra but helium is clearly detected.   The spectra of SN~Ic lack features of both hydrogen or helium.  This sequence is usually interpreted as evidence for progressive stripping of hydrogen and helium from the outer atmosphere of the progenitor star.  The stripping may be due to stellar winds or mass transfer to a companion star.  SN~Ib and SN~Ic are often described as ``stripped-envelope core-collapse" (SECC) events.

Type~IIb supernovae may form a separate class or they may be part of a continuum of related SN.  The spectra of SN~IIb display strong hydrogen features at early phases with no evidence for helium.  He features appear after about two weeks and they become stronger with time as the H features weaken rapidly.  The weakness of H features in SN~IIb suggests that they are related to the SECC group.  SN~IIb have been proposed as an intermediate step between SN~II and SN~Ib \citep{filippenko93, Nomoto93, Woosley94}.  For a review of the relationship between SN classifications and progenitor structure, see \citet{Nomoto95}.

There are two proposed progenitor channels for SN~IIb: an isolated star ($\gtrsim 25$~M$_{\sun}$) that loses most of its hydrogen envelope through stellar winds \citep[e.g.,][]{Chiosi86} or a close binary system where mass transfer strips most of the hydrogen from the outer layers of the progenitor star \citep[e.g.,][]{Podsiadlowski93}.  The progenitor in the mass-transfer model may be relatively low mass and compact, similar to a Wolf-Rayet star \citep{Dessart11}. 

The immediate post-explosion luminosity of SN~IIb is produced by a hot atmosphere that has been heated by the shock from the explosion.  The SN luminosity fades as the shocked atmosphere expands and cools.  SN~IIb from compact progenitors are expected to display relatively weak early emission from the shock heated atmosphere with the thermal powered luminosity declining rapidly (a few days) after the explosion \citep{Chevalier10}.  This behavior was observed in SN~2008ax \citep{Pastorello08, Roming09, Chornock11, Taubenberger11}.  On the other hand, SN~IIb with more mass and extended radii in their hydrogen envelopes are expected to have longer cooling times such as $\sim$20 days observed for SN~IIb~1993J \citep{Schmidt93, Wheeler93, Lewis94, Richmond94, Richmond96}.  A decline in the observed brightness of SN~2011dh is reported from day 1 to day 3 which is consistent with this cooling \citep{Maund11,Arcavi11,Bersten12,Tsvetkov12}.   

A second luminosity source is provided by radioactivity in the SN core.  The contribution from the core increases with time and within a few days it is equal to the declining thermal luminosity.  From that minimum, the SN brightness increases until it reaches the second, or radiation peak.   An initial thermal maximum, a decline to a minimum and subsequent rise to a second maximum have been observed in the light curves of several SECC SN: SN~IIb~1993J  \citep{Schmidt93, Wheeler93, Lewis94, Richmond94, Richmond96}, SN~IIb~2008ax \citep{Pastorello08, Roming09, Chornock11, Taubenberger11}, SN~IIb~2011fu \citep{Kumar12},  SN~Ib~2008D \citep{Modjaz09} and the peculiar SN~II~1987A \citep{Arnett89}.

The first spectrum of SN~2011dh was obtained by \citet{Silverman11} less than three days after discovery.  The presence of strong Balmer lines and the absence of helium led them to classify SN~2011dh as a SN~II.   Other reports noted similarities between the early spectra of SN~2011dh and Type~IIb SN~1993J without identifying He features \citep{Maund11, Arcavi11}.  \citet{Marion11} confirmed the Type~IIb classification by identifying \he\ lines in NIR spectra obtained 16 days after the explosion (\about 3 days prior to the \emph{B}-band peak).  The final version of \citet{Maund11} discusses \he\ lines in optical spectra obtained 20, 30 and 40 days after the explosion.  

Some authors find evidence for a compact progenitor of SN~2011dh.  \citet{VanDyk11,Prieto11,Arcavi11,Soderberg12,Krauss12} and \citet{Bietenholz12} use optical, radio, and x-ray observations and theory to suggest that the progenitor of SN~2011dh had a radius of order $1 R_{\sun}$.   \citet{Horesh12} derive an intermediate pre-explosion radius and suggest that a continuum of sizes for SN~IIb progenitors exists rather than a bimodal distribution consisting only of compact or extended stars.

Other authors conclude that the progenitor of SN~2011dh had an extended radius such as found in a supergiant type star.  The location of SN~2011dh was imaged by the \emph{HST} in multiple filters prior to the explosion and a yellow supergiant star has been identified near the position of the explosion.  An active debate in the literature discusses whether or not this star is a plausible progenitor.  Observational evidence in support of the yellow supergiant as the progenitor for SN~2011dh is presented by \citet{Maund11} and \citet{Murphy11}.   Theoretical support for the interpretation that the progenitor had an extended radius ($\ge 200 R_{\sun}$) is provided by \citet{Bersten12}, \citet{Benvenuto12} and \citet{Ergon13}.  

Almost 700 days after the explosion, \citet{VanDyk13} observed the location of SN~2011dh with the \emph{HST}/WFC3 and determined that the yellow supergiant star in that area has disappeared.  Their result suggests that the yellow supergiant was correctly identified as the progenitor and theoretical explanations can now focus on much more specific physical models.

Here, we report high cadence observations of SN~2011dh obtained between 4 and 34 days after the estimated date of explosion (May 31.5 UT).   Both photometric and spectroscopic data cover a wide wavelength range from 2,000~\AA\ in the ultraviolet (UV) to 2.4 ~\mum\ in the near-infrared (NIR).    The detailed observations of SN~2011dh presented here will help constrain theoretical models of SN~IIb and contribute to an improved understanding of SN~IIb and all stripped-envelope core-collapse supernovae.

The chemical structure of SN~2011dh is explored by measuring line-profiles and velocities for absorption features in the spectra.   Velocity measurements reveal the relative locations of line forming regions in radial space and the changing strengths of the lines with time provide clues about the relative abundances.   Observations with the Space Telescope Imaging Spectrograph (STIS) spectrograph on the \emph{Hubble Space Telescope} (\emph{HST}) and the \emph{Swift} U-grism provide rare UV spectroscopy of a SN~IIb and extend the range of our spectroscopic analysis.

Light curves (LC) are presented for twelve passbands: \emph{UVW2,UVM2,UVW1,U,u',B,V,r',i',J,H and K}.  We construct a bolometric light curve for SN~2011dh and compare the result to a bolometric light curve for SN~2011dh reported by \citet{Ergon13}.  We also compare the bolometric light curves of SN~IIb~1993J \citep{Richmond96} and SN~IIb~2008ax \citep{Pastorello08}.  The light curves for these SN~IIb display many similarities after the initial period of influence from the thermal luminosity.

The data acquisition and reduction details are presented in Section~\ref{obssec}.  Spectral features are identified and discussed in Section~\ref{specsec}.  Velocity measurements are described in Section~\ref{vsec} including a discussion of the gap between \h\ and \he\ velocities.  Synthetic spectra from SYNOW are used to confirm line identifications and to discuss physical properties in \S~\ref{synow}.  Multi-band light curves and color evolution are presented and discussed in \S~\ref{lcsec}.  The bolometric LC is described in \S~\ref{bolsec} and compared to other bolometric LC for SN~IIb.   Temperature estimates are discussed in Section~\ref{temps}.   A summary and conclusions are presented in \S~\ref{results}.

\section{The Observations}
\label{obssec}

The Whirlpool Galaxy (M~51) is a lovely galaxy that is frequently imaged by professionals and amateurs.  SN~2011dh was not detected in an image obtained by the Palomar Transient Factory (PTF) on May 31.275 to a limiting magnitude of $g=21.44$.  The first detection of SN~2011dh was by amateur A.~Riou on May 31.893 (UT) and there were several independent discoveries soon thereafter \citep{Reiland11}.  The PTF also made an independent discovery on June 1.191 and many early sources reference the supernova with the PTF name: PTF11eon \citep{Arcavi11}.  For this paper, we adopt an explosion date of May 31.5 UT (MJD 55712.5) and we express the times of all observations in days relative to the explosion.  Uncertainties of a few hours for the time of $t_0$ do not affect the presentation of our results or our conclusions.  

Distance measurements for M~51 have a significant scatter.  \citet{Tonry01} use surface brightness fluctuations (SBF) to determine $D = 7.7 \pm0.9$~Mpc.  However M~51 is a large spiral galaxy and there may be significant dust present that would compromise the effectiveness of the SBF method.  The \citet{Tonry01} result matches the value of $D = 7.7 \pm1.3$~Mpc found in Nearby Galaxies Catalogue that was determined using the Tully-Fisher method \citep{Tully88}.  A higher value for $D$ is found by  \citet{Vinko12} who use the expanding photosphere method (EPM) to measure the distance to M~51 as $D = 8.4 \pm0.7$~Mpc.  \citet{feldmeir97} also measure $D = 8.4 \pm0.6$~Mpc using a planetary nebula luminosity function.   

For this paper, we use the mean value of these four measurements for calculations that include distance ($D = 8.05$~Mpc).  We note that the maximum difference of $\pm0.35$~Mpc between our adopted value for $D$ and any of the referenced measurements introduces only $\pm0.1$~mag uncertainty in the luminosity measurements.  Such a small difference has no significant effect on our results.  We use a heliocentric velocity of 600 \kms\ (NED; \citet{Rush96}) to correct wavelength measurements to the rest frame.

Extinction from the host is low.  \citet{Vinko12} and \citet{Arcavi11} do not detect Na D lines at the redshift of M~51 in high-resolution spectra.  \citet{Arcavi11} set an upper limit on extinction from the host at $E(B-V) < 0.05$~mag.  \citet{Ritchey12} identify weak absorption components from Na D and \ca\ H\&K in high-dispersion spectra but they conclude that the overall weakness of the Na D detection confirms a low foreground and host galaxy extinction.   The \emph{HST}/STIS spectrum in our sample shows interstellar \mg\ ($\lambda 2795$) absorption with an equivalent width of approximately one~\AA, which is consistent with the assumption of low extinction.  

Galactic extinction is $E(B-V) = 0.035$~mag \citep{Schlegel98}, so that the \emph{V}-band extinction correction is \about 0.1 mag.  We correct for Milky Way extinction when constructing the bolometric luminosity but in the figures and tables we present the measured values. 

\subsection{Spectroscopy}

\begin{figure}[t]
\center
\includegraphics[width=0.5\textwidth]{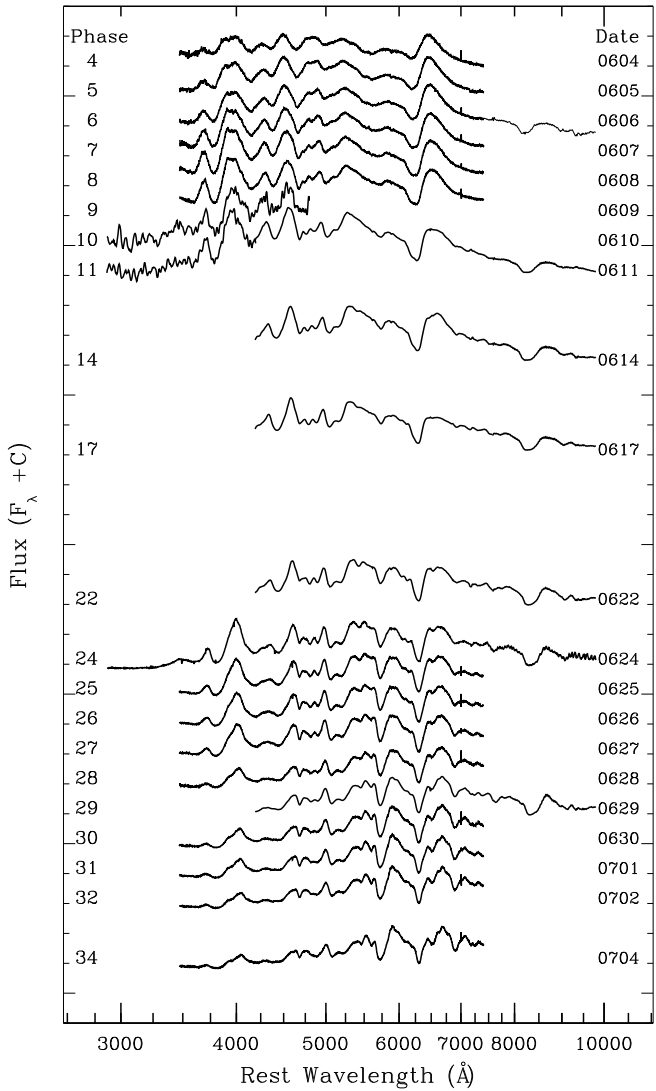}
\caption[]{UV and Optical spectra from SN~2011dh from 4 to 34 days after the explosion.  Data were obtained with the \emph{HST}/STIS (day 24, 2,800--10,000 \AA), Tillinghast/FAST (3,400--7,400 \AA), HET/LRS (4,200-10,000 \AA) and \emph{Swift}/U-grism (days 10 and 11, 2,800--4,600 \AA).   The phase with respect to the explosion is displayed to the left of each spectrum and the dates of observation are listed to the right.  Spectral features are discussed in \S~\ref{specsec}.  Observing details are listed in Table~\ref{spectbl}. \label{po11dh}}
\end{figure}

\begin{figure}[t]
\center
\includegraphics[width=0.5\textwidth]{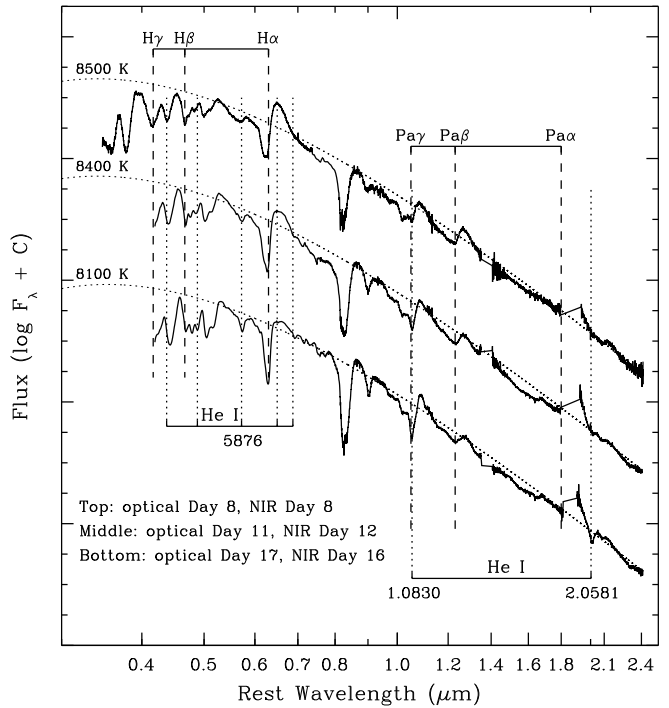}
\caption[]{Optical and NIR spectra of SN~2011dh from 0.32--2.4~\mum\ obtained 8 to 17 days after the explosion.   \h\ is strong at these early phases and Balmer and Paschen lines ($\alpha, \beta$ and $\gamma$) are marked at 12,000~\kms.   Several \he\ lines are marked at 8,000~\kms\ with the NIR $\lambda\lambda$1.0830, 2.0581 and the optical $\lambda$5876 lines labeled.  \he\ is not detected on day 8 (top).  The combined evidence of the labeled \he\ features suggests that \he\ is present on days 11 and 12 (middle).  \he\ features are obvious on day 16 (bottom).  The dotted lines are blackbody temperature curves. \label{poir}}
\end{figure}

Figure~\ref{po11dh} shows twenty-six spectra of SN~2011dh obtained between 4 and 34 days after the explosion.  The sample includes optical spectra (3,480--7,420~\AA) obtained at the F. ~L. \ Whipple Observatory (FLWO) 1.5-m Tillinghast telescope using the FAST spectrograph \citep{fabricant98} on days 4--9, 24--28, 30--32, and 34 with respect to the time of explosion.   FLWO/FAST data are reduced using a combination of standard IRAF and custom IDL procedures \citep{Matheson05}.  

Additional optical spectra (4,200--10,100~\AA) were obtained on days 6, 11, 14, 17, 22, and 29 with the 9.2m Hobby-Eberly Telescope (HET; \citet{Ramsey98}) at the McDonald Observatory using the Marcario Low-Resolution Spectrograph \citep{Hill98}.  HET/LRS spectra are reduced with standard IRAF procedures.  On some of the nights, two spectra were obtained that cover different wavelength regions.  In those cases, the spectra were combined to form a single spectrum for each phase.  The observational details are given in Table~\ref{spectbl}.

Low ($R \approx 200$, $\lambda = 0.65 - 2.5$ \mum) and medium ($R \approx 1200$, $\lambda = 0.80 - 2.4$ \mum) resolution NIR spectra were obtained on days 8, 12, and 16 with the 3 meter telescope at the NASA Infrared Telescope Facility (IRTF) using the SpeX medium-resolution spectrograph \citep{Rayner03}.  Figure~\ref{poir} displays the NIR spectra with nearly contemporaneous optical spectra.  IRTF data are reduced using a package of IDL routines specifically designed for the reduction of SpeX data (Spextool v. 3.4; \citet{Cushing04} 2004).  

Low-resolution UV spectra were obtained by \emph{Swift} with the UVOT U-grism (2,000--4,600~\AA) on days 10 and 11.  The data were downloaded from the \emph{Swift} archive\footnote{\rm http://heasarc.gsfc.nasa.gov/docs/swift/archive/}.  \emph{Swift} data are analyzed within HEAsoft following the standard recipe in the ``UVOT User's Guide."\footnote{\rm http://heasarc.gsfc.nasa.gov/docs/swift/analysis/\\ UVOT\_swguide\_v2\_2.pdf}  The raw frames are converted into `DET' images by applying the \texttt{swiftxform} task after flatfielding and bad-pixel masking.  The final spectra are extracted and calibrated with the \texttt{uvotimgrism} task.  Wavelength and flux calibrations are corrected by matching the U-grism spectra with contemporaneous ground-based spectra covering 3,000--5,000~\AA. 

A spectrum of SN~2011dh was obtained on day 24 with STIS on the \emph{HST} under observing program GO-12540 (PI: R.~P. \ Kirshner).  Three STIS gratings were employed with the CCD detector G230LB (3,600 sec), G430L (800 sec), and G750L (350 sec) and the combined spectrum covers 2,160--10,230~\AA.  The \emph{HST} data are reduced using the STScI STIS pipeline.

\subsection{Photometry}
\label{photo}

\begin{figure}[t]
\center
\includegraphics[width=0.5\textwidth]{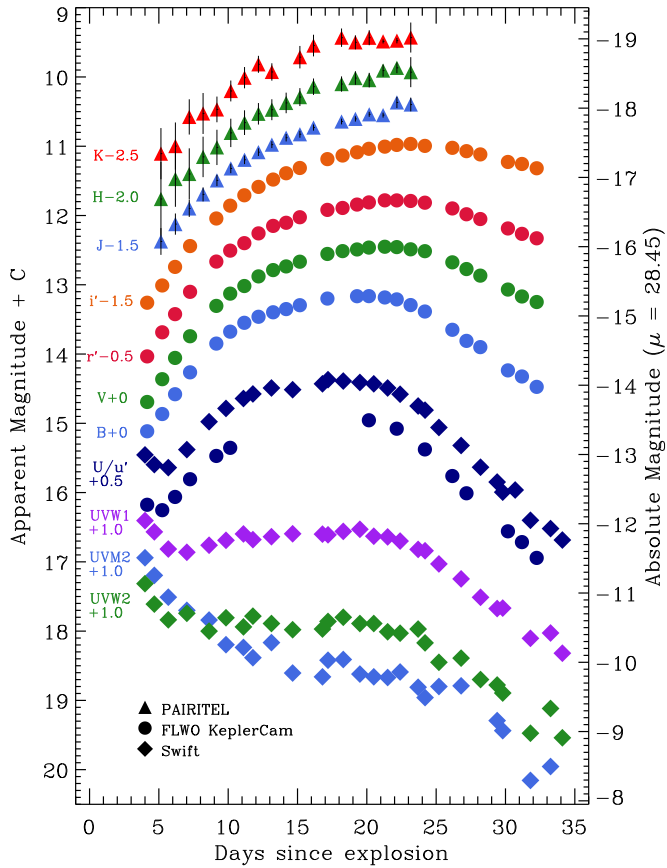}
\caption[]{UVOIR photometry of SN~2011dh in 12 filters obtained on days 4--34 by the \emph{Swift} satellite, the FLWO 1.5m with KeplerCam and the FLWO 1.3m with PAIRITEL.  Uncertainties are smaller than the symbols used for plotting except where indicated for some of the PAIRITEL data.  The \emph{Swift} \emph{U}-band and the KeplerCam \emph{u'}-band are plotted with the same offset and the light curves are different by about 0.5 mag.  We fit the data for each filter with a polynomial and list the peak magnitudes, dates of peak and the central wavelengths in Table~\ref{peaktbl}.  SN~2011dh had not reached maximum in the NIR before the final $JHK_{s}$ observations on day 24.  \label{lc}}
\end{figure}

\begin{figure}[t]
\center
\includegraphics[width=0.44\textwidth]{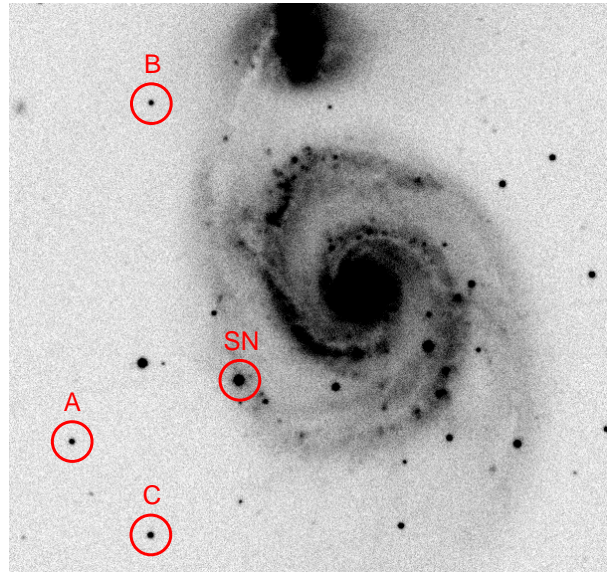}
\caption[]{M~51, SN~2011dh and local comparison stars used for differential photometry on non-photometric nights. (See Table~\ref{localcomp} and \S~\ref{photo}.) \label{fc}}
\end{figure}

Photometry of SN~2011dh was obtained in twelve filters and the light curves are displayed in Figure~\ref{lc}.  Table~\ref{peaktbl} gives the peak magnitudes and dates of peak for each passband.  The values were determined by fitting a parabola to the data.  The central wavelengths for each filter are also in the table.

SN~2011dh was observed at the FLWO with the 1.2-m telescope and the KeplerCam instrument in the \emph{uBVri} bands (Table~\ref{keptbl}).  The KeplerCam data are reduced using IRAF and IDL procedures as described in \citet{Hicken07}.  These data have not been s-corrected and no host-galaxy subtraction was performed since the SN was well separated from the galaxy center and other stars.  The  \emph{u'BVr'i'} instrumental magnitudes were measured with PSF-fitting on template-subtracted images using the methods described in \citet{Hicken12}.

Transformation to the standard photometric system was performed using local comparison stars around the SN in the same field-of-view. The linear transformation equations were calibrated using \citet{Landolt92} standards for \emph{UBV} and \citet{Smith02} standards for \emph{r'}- and \emph{i'}-bands. For \emph{u'} we transformed the \citet{Landolt92} \emph{U}-band magnitudes to \emph{u'} via the equation $u' = U + 0.854$ mag \citep{Chonis08}. The zero-points of the transformations were determined on 5 photometric nights.  The zero-points for images obtained on non-photometric nights were determined by differential photometry using local comparison stars as anchors (see Table~\ref{localcomp} and Figure~\ref{fc}). Note that our \emph{BV} data are in Vega-magnitudes, while the \emph{u'r'i'} data are in AB-magnitudes. 

The \emph{r'}- and \emph{i'}-band measurements of the field stars were checked against their magnitudes in the \emph{SDSS DR10} database.   The values were in agreement to \about 0.02 mag.  The same level of agreement was found when comparing the common field stars with the measurements by \citet{Arcavi11} and \citet{Tsvetkov12}.   However, systematic offsets of 0.04--0.15 mag exist between the FLWO \emph{u'BVr'i'} data of SN~2011dh and the measurements reported by \citet{Arcavi11} and \citet{Tsvetkov12}.  The offsets are essentially constant for all phases and have values of: \emph{u'}$= -0.05$ mag,  \emph{B}$= +0.15$ mag, \emph{V}$= +0.11$ mag, \emph{r'}$= -0.04$ mag and \emph{i'}$= +0.08$ mag.  Accounting for these offsets,  the FLWO light curves match \citet{Arcavi11} (\emph{i'}),\citet{Tsvetkov12} (\emph{u'BVr'}) and \citet{Sahu13} (\emph{UBVRI}) within $\pm 0.05$ mag.  The FLWO data in Table \ref{keptbl} and  Figure~\ref{lc} are our measurements and no offsets have been applied.

NIR images were obtained in the $J,H,K_{s}$ bands by the Peters Automated Infrared Imaging Telescope (PAIRITEL), a 1.3-m $f$/13.5 telescope located at the FLWO (Table~\ref{pteltbl}).  The data are processed into mosaics using the PAIRITEL Mosaic Pipeline version 3.6 implemented in python.  Details of PAIRITEL observations and reduction of supernova data can be found in \citet{friedman12}.

UVOT photometric data were downloaded from the \emph{Swift} archive (Table~\ref{swifttbl}). \emph{Swift} photometry is reduced with HEAsoft following standard procedures.  Individual frames were summed with the \texttt{uvotimsum} task, and magnitudes were determined via aperture photometry using the task \texttt{uvotsource} with a 5 arcsec radius aperture.  This sequence produces standard Johnson magnitudes in \emph{UBV} filters, and flight-system magnitudes in \emph{UVW2, UVM2} and \emph{UVW1}.  The transmission profiles of the \emph{UVW1} and \emph{UVW2} filters have extended red wings that permit flux contributions from beyond the intended wavelength regions.

\section{The Spectra}
\label{specsec}

The earliest spectrum in Figure~\ref{po11dh} was obtained on June 4.2 UT, which is less than four days after the estimated time of the explosion.  The final spectrum in our sample was obtained on July 4, which is 34 days after the explosion.  Observing details for each spectrum are given in Table~\ref{spectbl}.  The wavelengths of lines, features or spectra are expressed in Angstroms for wavelengths less than 8,000~\AA\ and in microns for wavelengths greater than or equal to 0.80~\mum.  When a discussion includes wavelengths that cross 8,000~\AA, all wavelengths are expressed in the same units.   The times of observation are expressed in whole days relative to the explosion.  As a point of reference, the \emph{B}-band maximum occurred about 20 days after the explosion.

Figure~\ref{poir} displays optical and NIR spectra from SN~2011dh covering the wavelength region 0.32--2.40~\mum.  The NIR data were obtained at the IRTF with SpeX on days 8, 12, and 16 while the corresponding optical data were obtained at the FLWO with FAST on day 8 and the HET with LRS on days 11 and 17.  Blackbody (BB) temperature curves are plotted as dotted lines in Figure~\ref{poir}.  The observed shapes of the OIR continua are fit best by BB curves for 8,500~K at 8 days, 8,400~K at 12 days, and 8,100~K at 16 days.  A more detailed discussion of temperature estimates can be found in \S~\ref{temps}.

\begin{figure}[t]
\center
\includegraphics[width=0.5\textwidth]{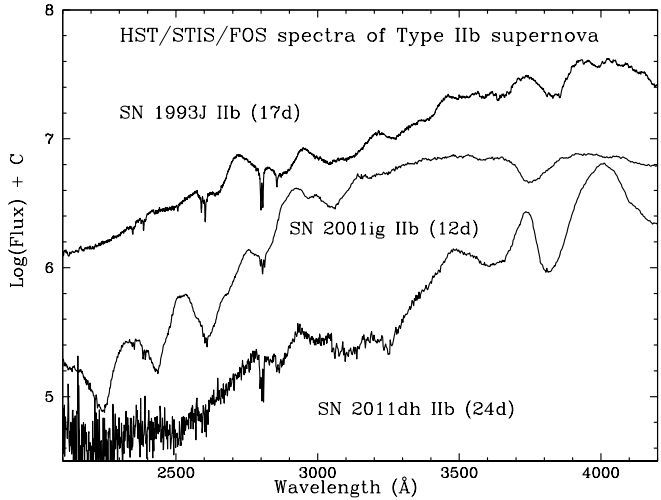}
\caption[]{\emph{HST}/STIS data from three Type~IIb~SN.  The spectra are displayed from 2,000--4,200 \AA\ to show detail in the UV region.  The SN~2011dh spectrum obtained on day 24 (bottom) reveals a reduced UV flux at wavelengths less than 3,300~\AA\ compared to spectra from SN~IIb~2001ig at day 12 (middle) and SN~IIb~1993J at day 17 (top).  We model this region with SYNOW (\S~\ref{synow}) and the results suggest that the UV suppression is due to line blanketing from \ti\ and \co.  \label{hstplot}}
\end{figure}

Figure~\ref{po11dh} shows that the \emph{HST}/STIS spectrum obtained on day 24 extends from 2,100--10,000~\AA, though the portion beyond 9,000~\AA\ suffers from fringing.  Features of \ca\ H\&K $\lambda$3945 and the \ca\ infrared triplet $\lambda$8579 are both observed in a single spectrum.  At other phases it is necessary to combine spectra from multiple sources to cover this range. The optical portion of the STIS spectrum is in good accord with a FAST spectrum obtained on the same date and with the HET spectrum obtained on day 22.

A unique part of the \emph{HST} spectrum is the UV coverage.  Figure~\ref{hstplot} shows the 2,000--4,200 \AA\ region of the \emph{HST}/STIS spectrum from SN~2011dh on day 24, plotted with STIS spectra of SN~IIb~1993J \citep{Jeffery94} and 2001ig at comparable phases.  The flux from SN~2011dh is clearly suppressed below 3,300 \AA\ relative to the other SN~IIb.  \fe, \ti, and \ion{Cr}{2} are expected to be strong contributors to the UV spectrum and these lines can create a line blanketing effect that shifts flux from the UV to longer wavelengths.  In \S~\ref{synow} we analyze the UV region of this spectrum using model spectra from SYNOW.  The spectrum of SN~2001ig is from the \emph{HST} archives and has not been previously published.

\subsection{Hydrogen Features}
\label{hsec}

\begin{figure}[t]
\center
\includegraphics[width=0.5\textwidth]{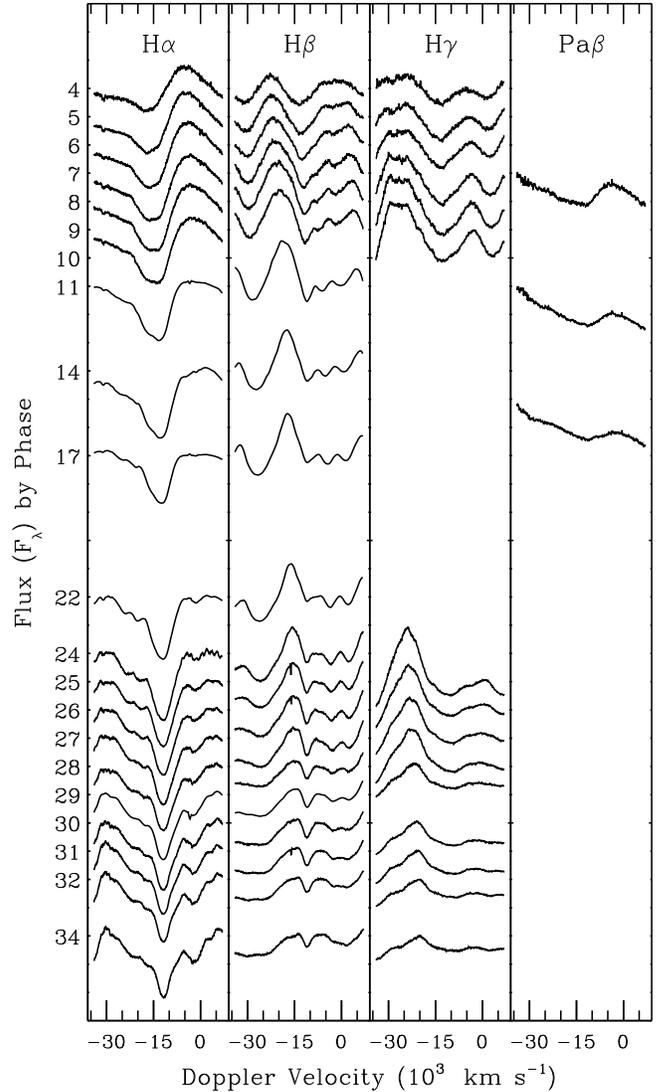}
\caption[]{Hydrogen features by phase in velocity space.   All \h\ features are strongest in the first spectra and they become weaker and narrower with time.  The \ha\ velocity is 15,400 \kms\ on day 4 and declines to about 12,500 \kms\ on day 14.   From day 14 to 34 the \ha\ velocity remains nearly constant at about 12,000 \kms.   Section~\ref{hsec} describes distortions and blends that affect each \h\ line as H weakens and other lines become stronger.  Flux values for Pa$\beta$ are multiplied by 5 to facilitate comparison.  Velocities are listed in Table~\ref{vtbl1}.  \label{pvh}}
\end{figure}

\citet{Silverman11} reported that a Keck I+LRIS spectrum of SN~2011dh obtained on day 3 revealed P~Cygni profiles for hydrogen lines that are characteristic of Type II supernovae.  This spectrum also appears in \citet{Arcavi11}.  The data in our sample begin the following day.  Figure~\ref{pvh} displays the strongest \h\ features by phase in velocity space.  The figure shows that \h\ features are strongest in the earliest observations and they become progressively weaker and narrower with time.  \ha\ is the only \h\ line that continues to form a significant absorption by the end of these observations, 34 days after the explosion.  \citet{Maund11} report that the transition from H dominated spectra to He is nearly complete 40 days after the explosion. 

In the first two weeks after the explosion, \ha\ line profiles show possible evidence for two separate absorption components.  At day four, the \ha\ absorption feature can be fit by two Gaussians with minima that correspond to velocities of 17,300 and 15,400~\kms.  \citet{Silverman11} reported the \ha\ velocity on day 3 to be 17,600~\kms\ which suggests that they measured the blue component.  Figure~\ref{pvh} shows the development of the \ha\ feature by phase.  The component on the red side of the double bottom becomes dominant after day 8 and by day 24 it provides the only minimum for the \ha\ line profile.  For consistency, we measure \ha\ velocities using the red component at all phases. 

The source of the blue component in the \ha\ features is ambiguous.  One possibility is a second H-rich region with a velocity about 2,000 \kms\ higher.  No comparable high-velocity (HV) components are found in the absorption features from other \h\ lines, but they could be disguised by blending.   SYNOW models suggest this is a possibility.  The addition of a second H region to SYNOW models can fit the observed ``double" \ha\ line profiles without producing strong HV features for other \h\ lines.    An outer H layer could also be created by interaction with circumstellar material as described by \citet{Chugai07}.

\si\ $\lambda$6355 is another possibility for the source of the blue absorption component in \ha, but that identification requires the \si\ velocities to be 5,000 \kms\ lower than \ca\ and 4,000 \kms\ lower than \fe\ at these early phases. Those velocities seem unlikely, although SYNOW models can produce features that match the observed \ha\ line profiles using \si\ for the blue component.  

Figure~\ref{pvh} shows that line profiles for most \h\ features are distorted by the influence of other lines that are gaining strength with time while \h\ is fading.   H$\beta$ velocities follow the general shape of the \ha\ velocity curve with values about 1500 \kms\ lower than \ha.  H$\beta$ is distorted on the red side by P~Cygni emission from a strong blend of \mg\ $\lambda$4481 and \fe\ $\lambda$4561.  Figure~\ref{pvh} shows how the H$\beta$ feature is shoved to longer wavelengths which results in lower measured velocities for H$\beta$.  After maximum brightness, \he\ $\lambda$4492 produces the same effect on H$\beta$.  

H$\gamma$ is squeezed by \ca\ H\&K emission on the blue side and by the \mg\ $\lambda$4481 and \fe\ $\lambda$4561 feature on the red side.  Our data do not cover the H$\gamma$ region between days 11--24 and it is not possible to identify H$\gamma$ in the spectra after day 24.

Pa$\alpha$ $\lambda$1.8751 is located in a wavelength region with high atmospheric opacity.  In Figure~\ref{poir}, we follow standard practice for displaying NIR spectra and omit the wavelength regions where the atmospheric transmission is less than 50\% (1.32--1.38 \mum\ and 1.79--1.88 \mum).  After those data have been removed, the only information we have about the Pa$\alpha$ line profile is the extreme red tail of the emission component.  

The \h\ line that is most free from blending is Pa$\beta$ $\lambda$1.2818.  Features from this line are displayed in the rightmost panel of Figure~\ref{pvh} and they show that the strength of the \h\ signal is clearly diminishing from day 8 through day 16.  Unfortunately our NIR spectra cover only a limited range of dates.
 
\subsection{Helium Features}
\label{hesec}

\begin{figure}[t]
\center
\includegraphics[width=0.5\textwidth]{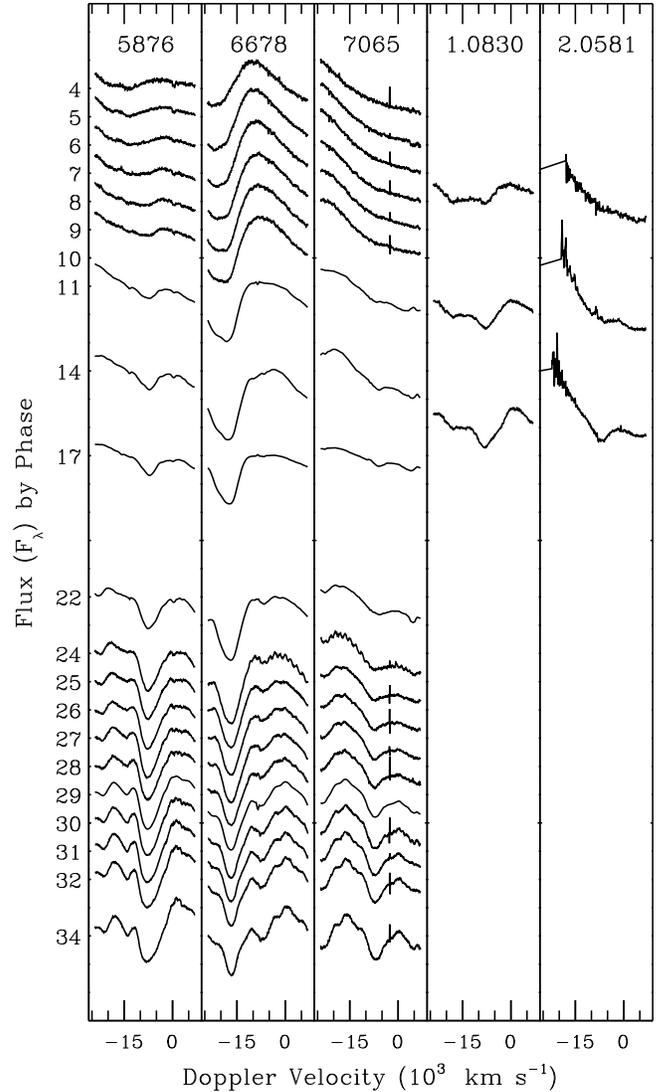}
\caption[]{\he\ features by phase in velocity space.  Helium is not detected until 11 days after the explosion when features of \he\ $\lambda$5876 and $\lambda$1.0830 (day 12) confirm the presence of \he\ (see also Figure~\ref{poir}).  \he\ features progressively strengthen through the time covered by these observations and velocities are nearly constant from day 14 to day 34.   Weaker lines such as $\lambda$6678 are not clearly detected until day 22.  Flux for the $\lambda$2.0581 feature is multiplied by 10 to facilitate comparison. Velocities are listed in Table~\ref{vtbl1}.  \label{pvhe}}
\end{figure}

The presence of helium in the spectra of SN~IIb differentiates them from other Type~II~SN but classification is only one reason to search for helium.  The development of \he\ features provides information about the transition in SN~IIb from a hydrogen dominated epoch soon after the explosion to the later phases dominated by helium.  NIR spectra provide the most reliable detections of \he\ features in early observations of SN because optical \he\ lines other than $\lambda$5876 are weak or blended.  Figure~\ref{poir} shows that \he\ lines at optical wavelengths are difficult to identify in the early data (see also the discussions of \he\ lines in SN 2008ax by \citet{Chornock11} and by \citet{Taubenberger11}).  

Evidence from the $\lambda$1.0830 line in isolation is insufficient to confirm the presence of \he\ (beside the fact that it violates the admonition in the previous paragraph).  A broad absorption feature from about 1.02--1.06 \mum\ is found in the spectra of most core-collapse SN, including SN~Ic that never display clear evidence for \he.  Pa$\gamma$ $\lambda$1.0938, \he\ $\lambda$1.0830 and \fe\ $\lambda$1.0500 are usually the most prominent contributors to this feature.   Several other strong lines are found in this region and a combination of them is likely to help form the line-profile.  Good candidates include: \ion{C}{1} $\lambda$1.0603, \mg\ $\lambda$1.0927, \ion{Si}{1} $\lambda\lambda$1.0608,1.0962, \ion{S}{1} $\lambda$1.0457, \fe\ $\lambda$1.0863 and \ti\ $\lambda$1.0691, but it is often difficult to define specific minima and associate them with individual lines. 
  
\he\ lines are marked in Figure~\ref{poir} at 8,000~\kms.  Eight days after the explosion (the top spectrum) no \he\ features are detected.  At 11 and 12 days (middle), \he\ $\lambda\lambda$5876,1.0830 features are evident while $\lambda$ 2.0581 is weak.  Taken together, these features show that helium is present at 11 days after the explosion but separately the features would not be convincing detections due to blends with \ion{Na}{1} $\lambda$5892 and Pa$\gamma$ $\lambda$1.2818 respectively.   The spectra obtained on days 16 and 17 reveal unambiguous features of \he\ from $\lambda$5876 and both strong NIR lines $\lambda\lambda$ 1.0830,2.0581 as described in \citet{Marion11}.  

In Figure~\ref{poir}, Pa$\gamma$ and \he\ $\lambda$1.0830 are positioned within 100 \kms\ of each other due to the velocity difference between the blueshifts for \h\ and \he.  Pa$\gamma$ dominates the red component of the 1.02--1.06 \mum\ feature on day 8 and all \h\ lines are strong at this phase.  On day 12, the red end of the feature has developed more of a P~Cygni profile as \he\ $\lambda$1.0830 begins to influence the line profile but Pa$\gamma$ remains strong.  \fe\ $\lambda$1.0500 probably forms the absorption at the blue end of this feature at all phases. 

The absorption feature for \he\ $\lambda$2.0581 is pushed to the red by the strong emission component of the Pa$\alpha$ P~Cygni profile that is mostly hidden in the high opacity region that we omit from the spectra.  

Figure~\ref{pvhe} displays features for three optical and two NIR \he\ lines by phase in velocity space.  Other \he\ lines begin to emerge about day 17 which is near the time of \emph{B}-max, and all \he\ features become deeper and broader with time.  In the early spectra \he\ $\lambda$3889 is blended or obscured by \ca\ H\&K and \si\ $\lambda$3858.  After \emph{B}-max, \he\ $\lambda$3889 is strong enough to distort the \ca\ H\&K feature by extending it on the blue side, but accurate measurement of \he\ contribution is not possible.  \he\ $\lambda$4492 is blended with \fe\ $\lambda$4561 and \mg\ $\lambda$4481 and it is not possible to unravel the \he\ contribution from the blend. \he\ $\lambda$5016 is blended with \fe\ $\lambda$5018 at all phases covered by these data.   \he\ $\lambda$6678 begins to flatten the emission component of \ha\ on day 14 but contributions from this line are obscured by \ha\ emission until 22 days after the explosion.  \he\ $\lambda$7065 is weak in spectra from days 14 and 17 but because this line is relatively unblended it provides a reliable benchmark for \he\ velocities at all phases.  Table~\ref{vtbl1} shows that the velocities for all \he\ lines are within 1,000~\kms\ of \he\ $\lambda$7065 after day 24 when \he\ has become established.

\subsection{\ca\ and \fe\ Features}
\label{othersec}

 \ca\ and \fe\ are the only unambiguous identifications in the spectra of SN~2011dh other than \h\ and \he.

\ca\ is clearly detected from both the infrared triplet (IR3, $\lambda$0.8579) and \ca\ H\&K ($\lambda$3945).  These strong lines create broad absorption features with rounded bottoms, which makes it difficult to identify distinct absorption minima.  \ca\ does not appear to have separate contributions from two absorption regions as found in the early \ha\ features.  To measure the \ca\ features, we define a straight line continuum between the flux at specific wavelengths on both sides of each feature.  Because the bottom of the line profile is a broad curve, small changes in the position of the assumed continua due to noise or imperfect telluric removals can create differences up to 2,000 \kms\ in the positions of the minima of these features.   Consequently velocity uncertainties for \ca\ are greater than for velocities measured from narrow features.  By using a consistent measurement technique we provide a satisfactory description of the behavior of \ca\ velocities in the spectra from SN~2011dh. 

On day 6 and day 11, the IR3 profile is similar to the combined profile of HV and photospheric \ca\ that is often seen in early spectra from Type Ia supernovae \citep{Mazzali05}.  But it is not possible to clearly define separate minima in the IR3 feature and \ca\ H\&K displays no evidence for a HV component at these epochs.  After day 17 the profile of the \ca\ IR3 feature becomes asymmetrical and the blue side has a steeper gradient than the red side.  The \ca\ H\&K feature becomes stronger from the first observation on day 4 through day 11.  This is followed by a gap in our coverage of this wavelength region from day 12 to day 23.  On day 24 \ca\ H\&K is still strong but the feature declines after day 27 and is very weak by day 34.  \ca\ H\&K is likely to be blended with \si\ $\lambda$3858 and \he\ $\lambda$3889.  

\fe\ becomes stronger from the earliest observations through day 14 and then slowly declines.  The $\lambda$5169 line is most frequently used to represent \fe\ velocity in SN~IIb \citep{Taubenberger11}.  An absorption feature from this line is detected in all optical spectra in our sample.   Unlike \h\ and \he, for which the velocity remains nearly constant after day 17, the \fe\ velocity steadily declines through the entire time period covered by our observations.   \fe\ $\lambda$4561 blends with \mg\ $\lambda$4481 to create a strong absorption feature in the early spectra.  This feature becomes stronger but even less defined when \he\ $\lambda$4492 begins to contribute about day 17.   \fe\ $\lambda$5018 is distinct on days 4--14.  This line is subsequently blended with \he\ $\lambda$5016 and eventually obscured as the helium lines become stronger.  \fe\ $\lambda$1.0500 is present and measured in all of the NIR spectra.

\subsection{Other Features}

The following ions show hints of absorption features from some of their stronger lines.  We discuss the most likely identifications.

\paragraph{\oi} is not detected through day 17 but may be emerging in the day 22 and 29 spectra from the $\lambda7773$ \oi\ line.  The absorption feature found near 0.90~\mum\ in the NIR spectra on days 8, 12, and 16 is not primarily due to \oi\ $\lambda0.9264$ because the stronger \oi\ $\lambda7773$ line is not detected before day 22.

\paragraph{\mg} is likely to be present in the NIR spectra from the $\lambda$0.9227 line that contributes to the strong absorption near 0.9 \mum\ in a blend with \oi\ $\lambda0.9264$ and \si\ $\lambda$0.9413.   \mg\ $\lambda$1.0927 is a part of the broad absorption feature from 1.01--1.06 \mum\ but no individual lines are identified in this blend other than Pa$\gamma$, \he\ $\lambda$1.0830 and \fe\ $\lambda$1.0500.  \mg\ $\lambda$4481 definitely contributes to the absorption feature found near 4,350~\AA\ in a blend with \fe\ $\lambda$4561.  The minimum of this feature corresponds to less than 7,000 \kms\ for \mg, which is inconsistent with the velocities of other intermediate mass elements.  This suggests that the 4350~\AA\ absorption feature is dominated by \fe\ early and \he\ $\lambda$4492 after day 17.

\paragraph{\si} does not produce obvious absorption features.  \si\ $\lambda$6355 may contribute to the \ha\ double feature (\S~\ref{hsec}).  However the minima of the blue components correspond to \si\ $\lambda$6355 velocities of 8,500 \kms\ on day 4 and less than 8,000 \kms\ on days 8--10.  Those velocities are more than 4,000 \kms\ less than any other lines measured during these phases.  \si\ $\lambda$3858 may influence the blue side of the \ca\ H\&K feature but detection is not possible.  \si\ $\lambda$4130 is a candidate for the flat top and slight absorption in the \ca\ H\&K emission component found near 3950~\AA\ in the early spectra.  But the velocities required are greater than observed for \ha.  In the NIR, \si\ $\lambda$0.9413 may contribute to the 0.9~\mum\ feature in a blend with \mg\ $\lambda$0.9227.  The small absorption features in the \emph{H}-band at about 1.65 \mum\ on days 12 and 16 could be from \si\ $\lambda$1.6930. 
 
\paragraph{\ti\ and \co} are the likely source of the steep drop of the continuum flux below 3,300~\AA\ due to line blanketing (Figure \ref{hstplot}).  Unlike SN~2001ig, there are no distinct absorption features in this region that can be associated with individual lines but SYNOW model spectra (\S~\ref{synow}) require significant contributions from \ti\ and \co\ to match the reduction of the UV-continuum level observed in early spectra from SN~2011dh.

\begin{figure*}[t]
\center
\includegraphics[width=0.8\textwidth]{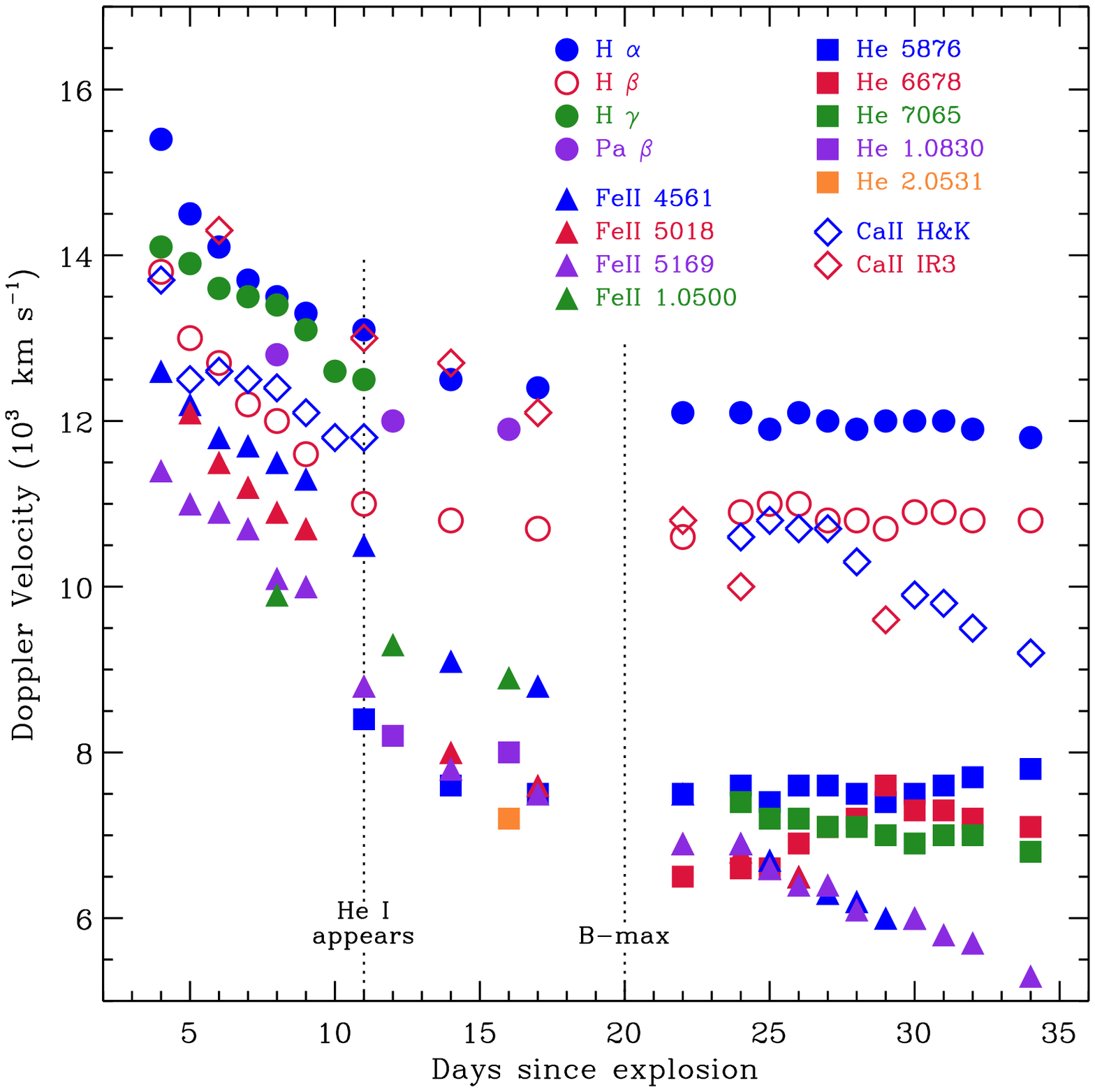}
\caption[width=1\textwidth]{Velocities of  \h, \he, \ca, and \fe\ plotted by phase.  Each ion has it's own symbol and each line has a different color.  All velocities decline rapidly until about day 11 when \he\ is first detected.   \h\ and \he\ velocities are nearly constant after \bmax, while \ca\ and \fe\ continue to decline.  The gap between the \h\ velocities (circles) and \he\ velocities (squares) is \about 4,000 \kms\  at all phases.  Open symbols indicate features for which there is increased uncertainty in the velocity measurements (see \S~\ref{hesec} and \ref{othersec}). Values used in the figure are listed in Tables~\ref{vtbl1} and \ref{vtbl2}.  \label{pvtbl}}
\end{figure*}

\section{Velocity Measurements}
\label{vsec}


\begin{figure}[t]
\center
\includegraphics[width=0.4\textwidth]{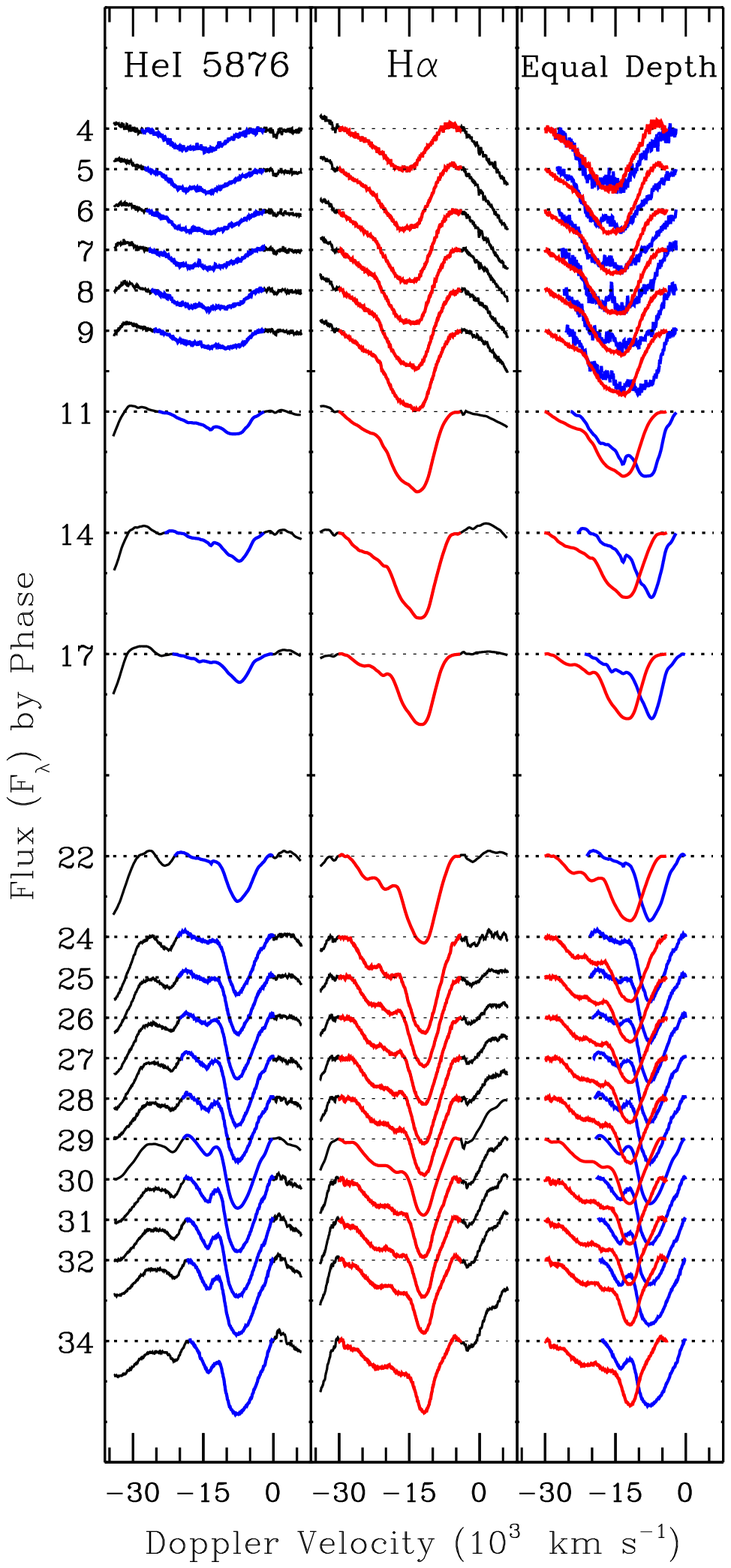}
\caption[]{\he\ $\lambda$5896 (left, blue) and \ha\ (middle, red) normalized to a flat continuum and plotted in velocity space.  The left and center panels have the same relative flux scaling.  \he\ becomes progressively stronger from day 11 through the end of our data.  \ha\ is strong early and becomes weaker and narrower with time.  The right panel shows \h\ and \he\ plotted together with the line depths for both features normalized to the same value for comparison.  By the time \he\ develops a clearly defined minimum at day 11, the \h\ and \he\ velocities are separated by about 4,000 \kms.  The velocity gap remains essentially constant from day 11 to day 34. \label{hhe}}
\end{figure}

Velocity measurements for \h, \he, \ca\ and \fe\ are plotted by phase in Figure~\ref{pvtbl} and they are listed in Tables~\ref{vtbl1} and \ref{vtbl2}.  The symbol shapes correspond to the ions: circles $=$ \h, squares $=$ \he, diamonds $=$ \ca\ and triangles $=$ \fe.  The different lines for each ion are represented by different colors.  Open symbols (H$\beta$ and \ca) indicate greater uncertainties in the velocity measurements.  

All velocities decline rapidly from the first detection through the time that \he\ appears (marked at 11 days on the figure.)  A more gradual rate of decline follows until the time of \bmax\ at day 20.  After maximum brightness, \h\ and \he\ velocities remain constant while \ca\ and \fe\ velocities continue to decline. 
 
Beginning about day 24, \ca\ velocities move away from the H-layer velocities toward the lower, He-layer velocities.  \fe\ velocities decline at a nearly constant rate throughout the entire period covered by these observations except for an abrupt drop from about day 9 to day 12.  This break may be due to an opacity effect because it occurs at about the same time that we begin to see evidence for \he.  After day 22, \fe\ velocities do not become constant, like \h\ and \he, but they continue to decline through the end of our observations.  \fe\ velocities separate from \h\ velocities about two weeks earlier than \ca\ velocities make the same transition.  The relative velocities between ions we find for SN~2011dh are similar to those reported for SN~2008ax by \citet{Taubenberger11}.  

\subsection{Separate line forming Regions for H and He}
\label{hhesec}

Figure~\ref{pvtbl} makes it easy to see that \h\ and \he\ velocities in SN~2011dh maintain a nearly constant separation of about 4,000 \kms\  during the period in which both ions are detected.  The detection of both \h\ and \he\ at the same phase but at different velocities was discussed by \citet{Branch02} for SN~Ib.  They use SYNOW models with a detached hydrogen component at 13,000 \kms\ and a helium region at 9,000 \kms\ to fit spectra of SN~Ib obtained near \bmax.  Models for SN~IIb by \citet{Dessart11} match our observations of the relative strengths of H and He features by phase.  They also predict that hydrogen velocities will always exceed helium velocities.   Previous observations of separated velocities for hydrogen and helium have been reported for SN~IIb~2008ax \citep{Chornock11} and SN~IIb/Ib~2011ei \citep{Milisavljevic13}.  

Figure \ref{hhe} presents another way to view the velocity relationship between the H and He regions.  \he\ $\lambda$5876 (left column, blue) and \ha\ (center, red) are plotted by phase and in velocity space with the same relative flux scale.  \he\ is absent or very weak at early times and it is blended with Na D.  The first detection of \he\ is on day 11 (\S~\ref{hesec}) and figure shows that  \he\ features become progressively stronger while \h\ features weaken.

The right panel shows the \h\ and \he\ features plotted together with their line depths normalized to the same value.  Generally similar line profiles are observed for the first few days but this is a coincidence with the \ha\ profile that has a double minimum and the \he\ profile that is strongly blended with Na D and has been magnified by the normalization.  

The minima of the features from \ha\ and \he\ $\lambda$5876 are obviously separated from day 11 when \he\ first develops a defined minimum.   From day 11 to day 22 the velocities for both lines decline slightly but the separation between them remains essentially constant.  From day 22 to day 34 the velocities for both lines are nearly constant.

These velocity measurements provide evidence that H-lines are formed in a region that is expanding \about 4,000~\kms\ faster than the layer in which He-lines are produced.   This separation is consistent with a model in which the progenitor explodes inside a hydrogen-rich shell.  In the first days after the explosion, H-shell opacity conceals all of the material inside the shell.  Opacity in the shell region is diluted by expansion of the SN and when the H-layer becomes optically thin, the helium-rich regions below are exposed.  In SN~2011dh this happens about day 11.    \citet{Milisavljevic13} also use H and He velocities to suggest that the progenitor of SN~2011ei retained a hydrogen envelope at the time of explosion.

\subsection{Mass of the Hydrogen Envelope}
\label{mass}

Velocity measurements of \h\ and \he\ in Type IIb supernovae 2011dh, 2008ax and 1993J provide enough information to compare the relative masses of the high-velocity, hydrogen-rich shells that surrounded the progenitors.  From a basic formula for the optical depth of the hydrogen layer we derive a simple model for the mass of the hydrogen shell.  We assume that the first detection of \he\ features occurs when the surrounding H-rich layer becomes optically thin.   Therefore: 

\begin{equation} \tau_{H} = 1 = \kappa_T \times \rho \times dR \end{equation}
where $\kappa_T$ is the electron scattering opacity, $\rho$ is the density and $dR$ is the thickness of the H layer.  We also assume homologous expansion for the SN, so $R \sim v_H \times t_{He}$.  The density is taken to be constant and we make a substitution: $\rho \propto M/R^3$.  Then we obtain:

\begin{equation}  M_H \propto (v_H \times t_{He})^2  \end{equation}
where $v_H$ is the velocity of the outer edge of the hydrogen shell and $t_{He}$ is the time after the explosion when the He layer is first observed.   Thus $t_{He}$ represents the time when the optical depth in H envelope reaches $\tau = 1$.  

For all three of the SN~IIb, the blue edge of the \ha\ absorption feature at the time of the first \he\ measurements is \about 20,000 \kms.   If we accept $v_H$ as a constant, then the mass of the hydrogen-rich shell is directly proportional to the square of the time from explosion to helium detection.  

We have shown that for SN~2011dh, $t_{He} \approx$ 11 days, and from the literature we find for SN~2008ax, $t_{He} \approx$ 4 days \citep{Chornock11, Taubenberger11} and for SN~1993J, $t_{He} \approx$ 18 days \citep{Barbon95}.   These values for $t_{He}$ indicate that the mass of the hydrogen shell surrounding the progenitor of SN~2011dh must be between the masses of the shells for SN~2008ax and 1993J.   

Using Equation 2 we calculate:
\begin{displaymath}  M_H(SN~2011dh) \approx 8 \times M_H(SN~2008ax)  \end{displaymath}
\begin{displaymath}  M_H(SN~2011dh) \approx 0.3 \times M_H(SN~1993J).  \end{displaymath}

Supporting evidence is provided by the times required for the luminosity to decline from the initial shock heated maximum to the minimum between the thermal and radiation peaks.  As for the time of \he\ detection, the time of the LC minimum for SN~2011dh is between the other SN~IIb.  The time to minimum is \about 4 days for SN~2011dh, \about 1 day for SN~2008ax \citep{Chornock11, Taubenberger11} and \about 9 days for SN~1993J \citep{Richmond94, Richmond96}.

\section{The Model Spectra}
\label{synow}

\begin{figure}[t]
\center
\includegraphics[width=0.5\textwidth]{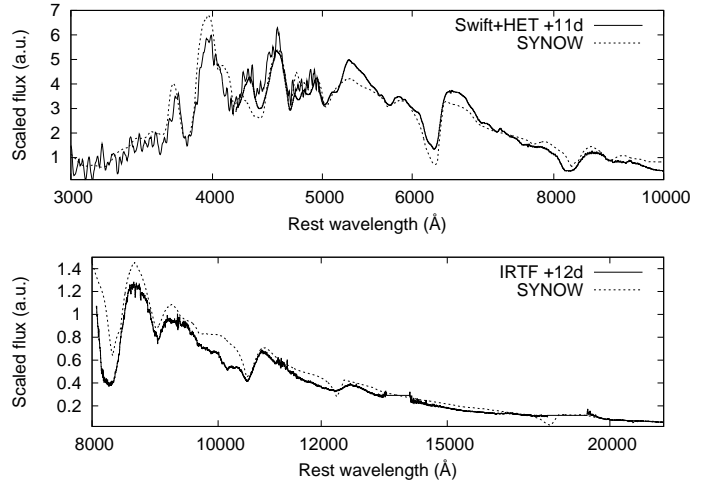}
\caption{SYNOW model spectrum (dashed line) and a combined \emph{Swift}/HET/IRTF spectrum (solid line) from days 11 and 12.   The top panel shows the UV-optical region (3,000--10,000~\AA), while the bottom panel shows the NIR (0.8-2.2 \mum) with an overlap of \about 2,000 \AA.  A detailed discussion of SYNOW modeling is given in \S\ \ref{synow}. \label{fig-synow1}}
\end{figure}

Figure~\ref{fig-synow1} shows the combined \emph{Swift}/HET/IRTF spectrum of SN~2011dh (solid line) obtained 11 and 12 days after the explosion plotted with a model spectrum covering 0.3--2.2~\mum\  (dashed line) created with the parameterized modeling code SYNOW \citep{Branch03}.  

The SYNOW model successfully fits the shape of the continuum and accounts for the presence of most spectral features.  The optical depth profile is assumed to be a function of velocity: $\tau = \tau (v_{\rm min}) \times (v / v_{\rm min})^{-n}$,  where $v_{\rm min}$ is the minimum velocity for a given ion.   The following ions (with their reference optical depths) are included in the model: \h\ (10), \he\ (0.2), \mg\ (0.7), \si\ (1), \ca\ (50), \ti\ (1), \fe\ (1), and \co\ (1).   

The best SYNOW fit to the complete OIR spectrum is achieved with a power-law index of $n=6$, a temperature of $T_{BB} = 9,000$~K, the velocity of H lines at $v_{\rm phot} = 12,000$~\kms\ and the velocities of all other lines at $v_{\rm phot} = 9,000$~\kms.  These are parameters for a simple line-fitting model and should not be considered physically precise.  But the SYNOW results clearly imply that the H line forming region is located at a higher expansion velocity than the region forming the other lines.  This is consistent with the observed separation of \h\ and \he\ velocities described in \S~\ref{hhesec}. 

\subsection{Early Evidence for \he}

\begin{figure}[t]
\center
\includegraphics[width=0.5\textwidth]{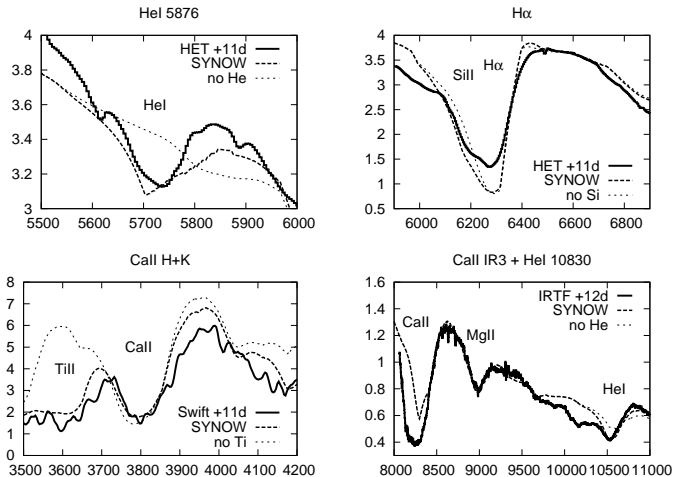}
\caption{SYNOW model (dashed line) and the OIR data from Figure~\ref{fig-synow1} with each panel zoomed in on specific spectral features: \he\ $\lambda$5876 (top left), \ha\ (top right), \ca\ H\&K (bottom left), and the \ca\ IR3 plus \he\ $\lambda$1.0830 (bottom right). A comparison spectrum from which specific ions have been removed is plotted as the dotted line in each panel: \he\ is omitted in the top left and lower right panels, \si\ is omitted in the upper right, and \ion{Ti}{2} and \ion{Co}{2} are omitted in the lower left. In each case the fit with the dashed line is better than the fit with the dotted line.  A detailed discussion of SYNOW modeling is given in \S\ \ref{synow}.  \label{fig-synow2}}
\end{figure}

SYNOW models can help determine when \he\ becomes present in the spectra of SN~2011dh.  \he\ detections on day 11 have been discussed and we seek confirmation from the models for these identifications.  Figure~\ref{fig-synow2} zooms in on individual features to facilitate a detailed comparison of the model to the data.  

The top left panel shows that the weak absorption feature near 5750~\AA\ (solid line) is most accurately modeled (dashed line) when \he\ is included with a low optical depth (0.2).  Note that the flux scale in this panel makes the feature appear stronger that it actually is relative to the other features in Figure~\ref{fig-synow2}.  The model does not reproduce the line-profile exactly, but it matches the size and location of the feature.  The dotted line shows that a model without \he\ generates a poor fit to the data.  The lower right panel of Figure~\ref{fig-synow2} shows that the feature observed near 10500 \AA\ in the day 12 data (solid line) is primarily due to Pa$\gamma$ (dotted line).  But the model fits the data better when \he\ is included (dashed line).  

Thus the SYNOW models suggest that \he\ is present 11 days after the explosion even though \he\ does not produce strong absorption features at this phase.  Model results are not detections but when they are consistent with the observational evidence (\S~\ref{hesec}) they reinforce the identifications of  features such as \he\ $\lambda\lambda$5876,1.0830 in the day 11 and 12 spectra.


\subsection{\ha\ Features with Two Components}

We use SYNOW to explore the double bottom found in the \ha\ profile of early spectra from SN~2011dh. The \ha\ feature is displayed in the upper right panel of Figure~\ref{fig-synow2}.  The data (solid line) show that the blue component on day 11 is weaker than found in earlier data but still forms a distinct notch in the blue side of the profile.   The dashed line is from a model that includes \si\ at the same velocity as \he\ and although it produces a deeper absorption than observed, it shows a notch near 6200~\AA\ that is similar in shape to the observed feature.  The dotted line is from a model without \si\ and it fails to reproduce the observed inflection.  

SYNOW can also produce a notch like the one found in the data when we include a high-velocity hydrogen line forming region at 15,000 \kms\ in addition to the H layer that forms the primary absorption at 12,000 \kms.  We model the HV region with a shallower optical depth profile (n= 5 instead of 6).  The model with HV hydrogen only generates a detectable blue component for \ha.  It does not generate similar HV notches for the other \h\ lines.  That result is consistent with the observations that do not find HV components (notches at 15,000 \kms) for other \h\ lines (Figure~\ref{pvh}).   

In this case SYNOW provides two plausible answers to the question about the source of the two-component line profile, but it does not give clear guidance to help us choose between them.

\subsection{\ca\ Features}

The bottom left panel of Figure~\ref{fig-synow2} shows that the \ca\ H\&K $\lambda$3945 feature (solid line) is successfully fit with a SYNOW model using a $n=6$ optical depth profile (dashed line).   But the same model does not fit the observed \ca\ infrared triplet (IR3, $\lambda$8579) seen in the bottom right of Figure~\ref{fig-synow2}.  The IR3 forms a stronger and more extended absorption than the $n=6$ model used to fit \ca\ H\&K.   The model can be improved with respect to the IR3 feature by using a shallower optical depth profile ($n=5$) or by inserting a HV \ca\ layer, but both of these solutions produce spectra that are incompatible with the observed \ca\ H\&K profile.  The unusual IR3 profile may be caused by line blending and not HV \ca\ but we are unable to reproduce the observations of \ca\ IR3 with any of the other ions expected to be found in Type II SN atmospheres \citep{Hatano99}.

\subsection{UV Flux Deficit}

We also use SYNOW models to explore the steep drop in the continuum flux of SN~2011dh at the blue end of the \emph{HST}/STIS spectrum compared to other SN~IIb (Figure~\ref{hstplot}).  This wavelength region is included in the lower left panel of Figure~\ref{fig-synow2}, but at this phase the data are from \emph{Swift}/U-grism observations.  The model spectrum plotted with a dashed line matches the data (solid line) reasonably well and contributions from both \ti\ and \co\ are required to fit the observed UV-continuum level.  The dotted line is produced by a model that omits both \ti\ and \co\ and it clearly has excess flux at wavelengths less than 3700 \AA.  Despite the suggestion from SYNOW that \ti\ and \co\ are present, individual absorption features from \ti\ and \co\ are not  identified in the \emph{Swift} data from days 10 and 11 or the \emph{HST} from day 24.

\section{The Light Curves}
\label{lcsec}

Figure~\ref{lc} displays light curves for SN~2011dh in 12 filters covering 4--34 days after the explosion.  For all figures and tables we express the times of observation in whole days relative to the estimated date of explosion.  Data were obtained with \emph{Swift} filters (\emph{UVW2}, \emph{UVM2}, \emph{UVW1} and \emph{U}), FLWO/KeplerCam filters (\emph{u', B, V, r'} and \emph{i'}) and FLWO/PAIRITEL filters ($J,H,K_{s}$).   The measurements for each filter are listed by phase in Tables \ref{keptbl}, \ref{pteltbl}, and \ref{swifttbl}.  The data for bands from \emph{UVW1} through \emph{i'} were fit with a polynomial to estimate the peak luminosities and the times of maximum brightness that are displayed in Table~\ref{peaktbl}.  

All references to ``peak" or ``maximum" in the following discussion refer to the observations in this sample that cover the second, or radiation powered peak.  We note that systematic offsets of 0.04--0.15 mag are found between the FLWO/KeplerCam \emph{u'BVr'i'} data and other published optical data (see \S~\ref{photo}).  The data presented here are our measurements and no offsets have been applied. 

Evidence for cooling from the initial shock-heated peak can be seen in the first few observations that show a decline in brightness for the \emph{U}-band and bluer passbands.  In our sample, minima are reached on day 5 for \emph{u'}, day 6 for \emph{U}-band and day 7 for \emph{UVW1}.  The UV light curves subsequently rise toward a second peak as radioactivity in the core begins to provide the luminosity.   At the phase of our first observations (4 days after the explosion) the light curve for the \emph{B}-band is at minimum and the light curves for all longer wavelength filters have passed their minima and are already rising toward the radiation peak.   An initial decline was also observed at optical wavelengths in data of SN~2011dh obtained prior to day 4 \citep{Maund11,Arcavi11,Bersten12}. 

The \emph{U}-band peak occurs on day 16 and as the filter wavelength increases the time to peak increases.  SN~2011dh reached a \emph{B}-band peak of 13.18~mag on day $20.0 \pm 0.5$ and a \emph{V}-band peak of 12.44~mag on day $20.6 \pm 0.5$.  The dates and magnitudes of the $JHK_{s}$ maxima are uncertain because we lack NIR photometry after day 24, although inspection of Figure~\ref{lc} suggests that the NIR light curves are within a day or two of their maxima on day 24.  

Figure~\ref{lc} shows that light curves for all bands experience a relatively steep initial rise in brightness with a noticeable change in slope between days 9--12 to a more gradual rate of increase.  This phase corresponds to the initial detection of \he.  From that phase through the time of maximum brightness the rate of increase is slower.  After the peak in each band the decline is almost linear through the end of our data on day 34.  This pattern is similar in all bands from \emph{UVW1} to \emph{H} with the phase of the changes occurring earlier at shorter wavelengths.   The LC of SN~2011dh presented by \citet{Sahu13} and \citet{Ergon13} also show this behavior.

The light curve for \emph{UVM2}, which is the shortest wavelength filter in our sample, declines steeply from the thermal peak through about day 15 but the subsequent radioactive diffusion is at too low a temperature to revive the light curve.  The luminosity in\emph{UVW2} is essentially flat from day 6 through day 23 which is the time of bolometric maximum.  \emph{UVW1} is the bluest filter to exhibit a radiation powered second peak.

The transmission curves of the \emph{Swift}/UVOT \emph{UVW1} and \emph{UVW2} filters have extended red tails so the measured flux in these bands includes a contribution from the red side that is beyond the intended wavelength region.  We have shown that SN~2011dh has a very low UV flux during the time of our observations, and it is likely that there are significant contributions from optical wavelength regions to the \emph{Swift} measurements in these filters.  This ``red leak" may contribute to the differences in timing of the minima and overall LC shape between the \emph{UVM2} band and the \emph{UVW1} and \emph{UVW2} bands.

\subsection{Photometric Colors}

\begin{figure}[t]
\center
\includegraphics[width=0.5\textwidth]{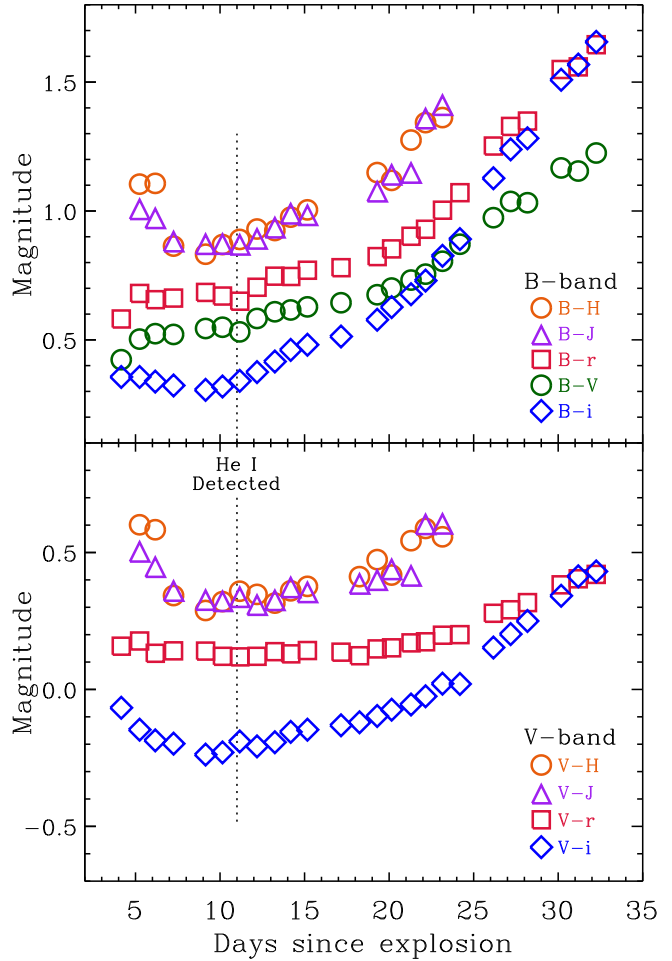}
\caption[]{B- and V-band colors for SN~2011dh. The optical minus NIR colors are redder at the earliest observations and decline rapidly through day 9 while optical minus optical colors are bluer in the initial observations and remain nearly constant over days 4--9.   All colors redden from day 10 through day 34, which indicates the atmosphere is cooling.  \label{colors}}
\end{figure}

\emph{B}- and \emph{V}-band colors for SN~2011dh are plotted by phase in Figure~\ref{colors}.  At day 4, the optical minus IR colors ($B-J, B-H, V-J, V-H$) are \about 0.5 mag redder than optical minus optical colors at this early phase.  The optical minus IR colors rapidly become bluer at about 0.035 mag/day through day 9 while the optical minus optical colors show little change through day 11.   

At about day 9, the optical minus IR colors reach a minimum and immediately start reddening, but with a slower rate of change.  The timing of these minima coincides with transparency in the H shell since \he\ is first detected on day 11. This is also the phase when measured velocities of most absorption features stop their rapid decline (Figure~\ref{pvtbl}).   From day 10 through the time of maximum brightness (\about day 21) all colors shift gradually to the red, indicating that the atmosphere is cooling.  After maximum, expansion continues to cool the SN and the rate of reddening increases.  The color evolution that we find for SN~2011dh agrees with the colors reported by \citet{Sahu13} and \citet{Ergon13}.

The $B-V$ colors in SN~2011dh is similar to that of SN~1993J with a period of reddening through about day 8 followed by a very gradual reddening for about two weeks and then a steeper increase \citep{Barbon95,Richmond96}.    $B-V$ in SN~2011dh is \about 0.2 mag greater from day 4 to day 22 but very close to $B-V$ in SN~1993J from day 22 through the end of our observations.  

From day 4 to day 20, $B-V$ development in SN~2008ax is very similar to the $B-i'$ behavior observed in SN~2011dh.  The color gets bluer from the initial measurement, reaches a minimum near 0.2 mag on day 12 and then begins to redden.  By day 22, $B-V$ in SN~2008ax is close to the same values found in SN~201dh and SN~1993J.  For both SN~1993J and 2008ax, $B-V$ stops increasing about day 40.

The $V-R$ color curve of SN~2011dh is generally similar to $V-R$ color behavior in SN~Ib and SN~Ic found by \citet{Drout10}.  These SN types are similar to SN~IIb in that they are core-collapse events with little or no hydrogen in the spectra.  The mean $V-R$ values from \citet{Drout10} display an early decline through day 12 (with respect to the explosion), a period of little change through day 20, and an increase through about day 36.  The $V-R$ curve in SN~2011dh is slightly flatter but $V-i'$ is a close match.  

\section{The Bolometric Light Curve}
\label{bolsec}

\begin{figure}[t]
\center
\includegraphics[width=0.5\textwidth]{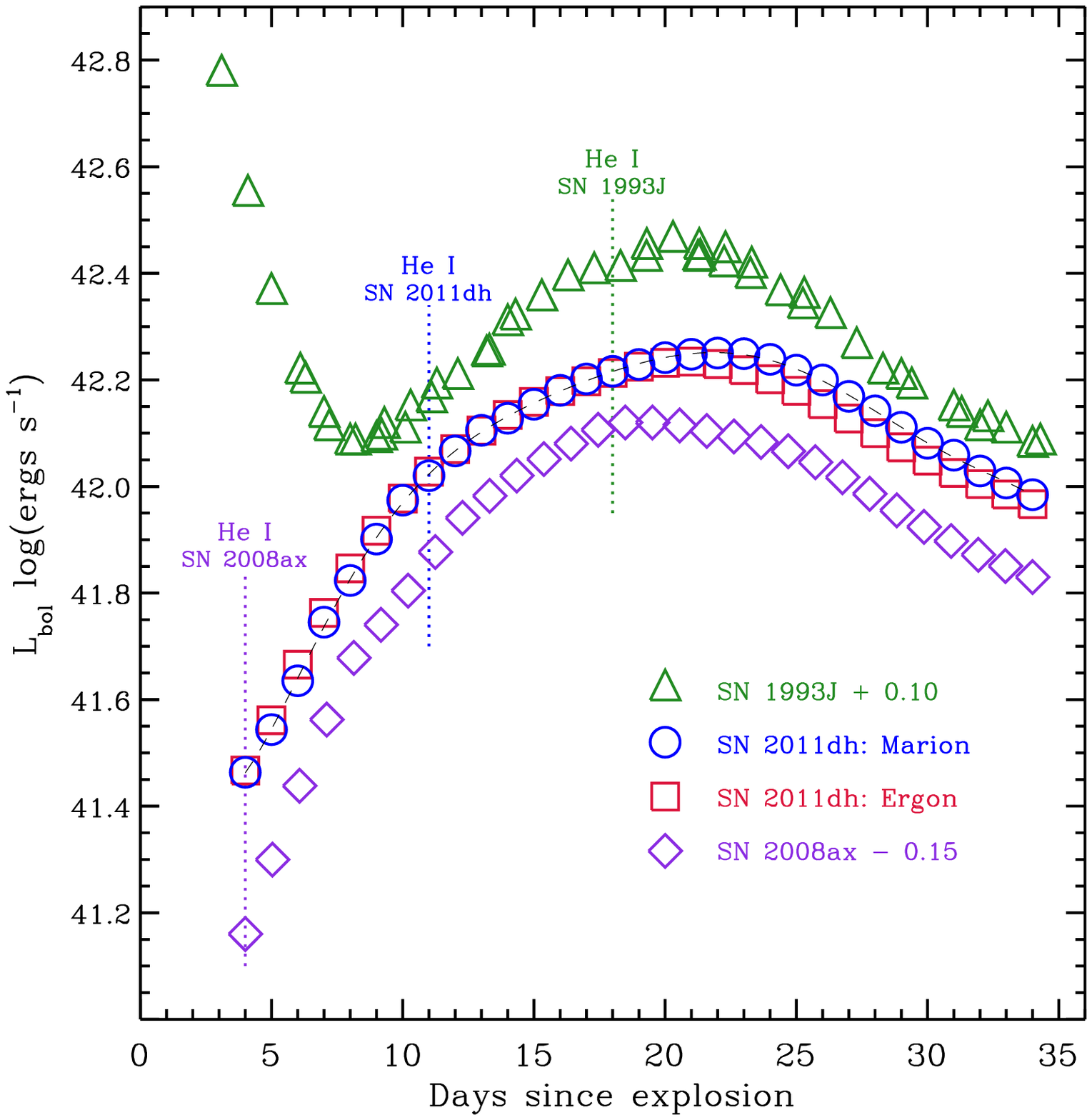}
\caption[]{Bolometric light curves of three SN~IIb.  SN~1993J \citep{Richmond94} is displayed as green triangles and is shifted up by 0.10 log(erg s$^{-1}$) for clarity.  SN~2011dh data from this paper is displayed with blue circles and SN~2011dh data from \citet{Ergon13} are shown as red squares.  SN~2008ax \citep{Pastorello09} is plotted with purple diamonds and the LC is shifted down by 0.15 log(erg s$^{-1}$).  The LC for the three SN~IIb LCs are very different for the first twelve days, but the maximum luminosities, times of maxima and decline rates are similar.     The phases at which \he\ was first detected in each of the SN~IIb are marked with the same colors as the symbols.\label{bol}}
\end{figure}

We construct a bolometric light curve for SN~2011dh from UVOIR data obtained by \emph{Swift}, KeplerCam and PAIRITEL.  The flux measurements are given by phase in Table~\ref{boltbl} and the LC is plotted in Figure~\ref{bol}.  A second bolometric LC for SN~2011dh, compiled by \citet{Ergon13} from a different data set, is also displayed.  Bolometric light curves for two other Type~IIb~SN are also displayed.  SN~1993J \citep{Richmond94} is shown with green triangles and 2008ax \citep{Pastorello08} is plotted with purple diamonds.  

The SEDs were sampled at the central wavelength of each observed passband.  Measurements were corrected for Galactic reddening $E(B-V) = 0.035$~mag using the total-to-selective extinction ratios for all filters given in \citet{Schlegel98}.   Magnitudes were converted to quasi-monochromatic flux densities using the central wavelengths for each filter as listed in Table~\ref{peaktbl}.  The flux was integrated by wavelength using a simple trapezoidal rule.  The UV contribution was determined by integrating the reddening-corrected Swift UV-fluxes for filters \emph{UVW2,UVM2} and \emph{UVW1}.   The missing far-UV fluxes were estimated by assuming zero flux at 1,000 \AA\ and approximating the SED with a straight line between 1,000 \AA\ and $\lambda_c$(\emph{UVW2}) = 2,030 \AA.  The missing mid-IR flux was approximated by integrating the Rayleigh-Jeans tail of a blackbody from the flux measured at the central wavelength of the reddest filter to infinity.  The total measurement uncertainties are \about 10\%.  

The maximum bolometric luminosity of $1.8 \times 10^{42}$ erg s$^{-1}$ was reached about 22 days after the explosion (Table~\ref{boltbl}).   NIR contributions to the total bolometric flux account for about 30\% of the total bolometric flux from day 4 through the bolometric maximum.   After maximum, the NIR fraction increases and it reaches 49\% on day 34.  The UV fraction of the total is 16\% on day 4, it declines rapidly to 5\% on day 9 and then to 1\% on day 29.    

Our results for the bolometric luminosity and fractional contributions of the UV and NIR agree with the results presented by \citet{Ergon13} and by \citet{Lyman13}.  Figure~\ref{bol} shows that the  bolometric LC for SN~2011dh from this paper and from \citet{Ergon13} are very close before the bolometric maximum.  After maximum there is a small divergence due to the way that the NIR contribution is calculated.  Our sample lacks NIR photometry after day 22.  From 23--34 days the reddest filter in our sample is \emph{i'}.  That means that the blackbody used to estimate the missing IR flux begins at 0.775 \mum.  \citet{Ergon13} include MIR data from \emph{Spitzer} and they calculate the missing IR contribution with a blackbody beginning at 4.5 \mum.  

The \citet{Richmond94} bolometric LC for SN~1993J in Figure~\ref{bol} is based on optical (\emph{UBVRI}) observations.  The missing contributions on both the blue and red sides are estimated by using the optical data to establish a temperature at each phase.  The amount of blackbody flux that would be emitted outside the \emph{UBVRI} window is added to the total from the optical measurements.  The bolometric light curve for SN~2008ax was produced by \citet{Pastorello08}.  They compiled several data sets of SN~1993J to construct a new UVOIR LC and then rescaled the \emph{UBVRI} data for SN~2008ax data using the same fractional NIR contribution estimated for SN 1993J.

The early LC from SN~IIb 2011dh, 2008ax and 1993J have very different shapes due to differences in the cooling times of the shock heated material surrounding each SN.  But the maximum bolometric luminosities, the times of maxima, the decline rates and the $B-V$ colors are similar for all three SN~IIb.  The near uniformity in the bolometric light curves after the first few days suggests that the hydrogen shells only affect the LCs during the early, cooling phase.  The similar bolometric luminosities suggest that the $^{56}$Ni yields must have been close to the same for each explosion.  These conclusions are all consistent with progenitors of SN~IIb having similar compositions and masses but they explode inside hydrogen envelopes that can vary significantly. 
 
\vspace{0.5cm}

\section{Temperatures}
\label{temps}

\begin{figure}[t]
\center
\includegraphics[width=0.5\textwidth]{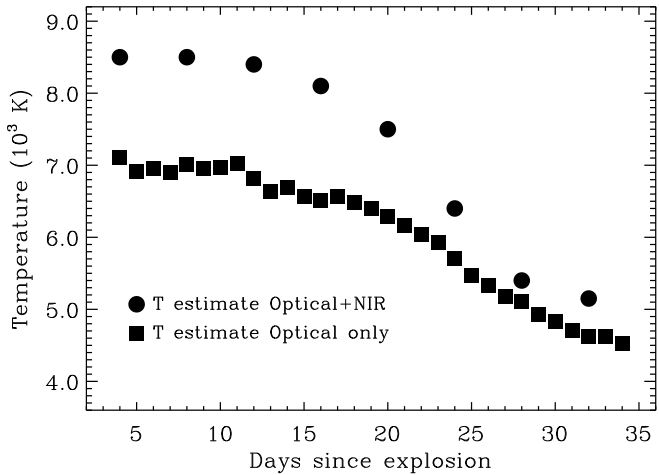}
\caption[]{Estimated temperature by phase for SN~2011dh from 4--34 days after the explosion.  The circles are temperature estimates based on blackbody fits that include NIR data.  The squares are temperature estimates made using optical data only.  \label{ptemp}}
\end{figure}

Figure~\ref{ptemp} shows estimated temperatures for SN~2011dh by phase.  The filled circles are temperature estimates obtained by fitting blackbody (BB) curves by eye to a full set of OIR SEDs from passbands \emph{BVr'i'JHK}.  The filled squares are temperature estimates using only optical SEDs (\emph{BVr'i'}) that are fit to BB curves by minimizing $\chi^{2}$.   Both methods suggest very little change in temperature from day 4 to day 12 that is consistent with nearly constant values for $B-V$ during this epoch.  The phase at which the temperature begins to decline is also when the H-layer becomes transparent and reveals the He-region below.

Figure~\ref{poir} shows BB curves fit to the OIR spectra obtained 8, 12, and 16 days after the explosion.  We find a very consistent relationship between BB temperature curves, the complete OIR spectra, and the SEDs for OIR passbands \emph{u'BVr'i'JHK}.   We do not plot SEDs on the already crowded Figure~\ref{poir}, but the temperature curves fit to OIR spectra pass directly through the SEDs for passbands \emph{Vi'JHK} at all three phases.  The circles in Figure~\ref{ptemp} represent the temperature of a BB curve that we chose by eye to fit the \emph{Vi'JHK} SEDs.  Line blending is less severe in the NIR than at UV and optical wavelengths, which puts an additional premium on the PAIRITEL data that are available through day 24.  After day 24, we replace the $JHK$ SEDs by integrating the Rayleigh-Jeans tail of a blackbody.

The SEDs for \emph{u', B} and \emph{r'} are less reliable.  For example, the \emph{r'}-band SED is complicated by the large \ha\ absorption feature in spectra of SN~2011dh.  BB curves that pass through the \emph{V}-band and \emph{i'}-band SEDs, are always a little below the \emph{r'}-band SED.  The suppressed continuum flux for SN~2011dh at less than 4,000 \AA\ means that a blackbody is not a realistic representation of data from SN~2011dh at these wavelengths.  That means that any method of estimating temperature with BB curves does not actually fit data to the peak of the BB curves.   

Temperature estimates made by fitting to a wider wavelength range that includes the NIR appear to be more effective than estimates made with optical data alone.  Our estimate for the day 4 temperature is 8,500~K which is \about 1,000~K higher than the temperature estimate for a day 3 optical spectrum provided by \citet{Arcavi11} and \about 700~K higher than the estimate of \citet{Bersten12}.    

\section{Summary and Conclusions}
\label{results}

Our results show details of chemical layering in SN~2011dh that have not been previously reported for SN~IIb.  We demonstrate the importance of NIR measurements for line identification, temperature estimates and bolometric estimates.    
  
A UV spectrum from \emph{HST}/STIS shows a significant deficit in continuum flux at wavelengths below 4,000 \AA\ when compared with other SN~IIb.  SYNOW model spectra are used to demonstrate that line blanketing from \ti\ and \co\ are responsible for the suppressed UV flux.  We  also use SYNOW models to investigate blended features and to support spectral evidence for the presence of \he\ in a spectrum obtained on day 11. 

Velocities of \h, \he, \ca, and \fe\ lines are measured from 4 days after the explosion through day 34.  All velocities are at maximum in the earliest observations and decline rapidly to about day 10.  The change in the decline rate coincides with the first detection of \he\ in the spectra.  \h\ and \he\ velocities become constant by day 20 while \ca\ and \fe\ velocities continue to decline.  The relative velocities between ions and the velocity decline rates found in SN~2011dh are similar to observations of SN~IIb~1993J and SN~IIb~2008ax. 

We show that early measurements of both \h\ and \he\ are important for understanding the transition in SN~IIb from the early phases when hydrogen dominates the spectra to later times when H fades and He becomes dominant.  We present the earliest secure identification of \he\ lines in spectra of SN~2011dh.  NIR spectra confirm that \he\ is present 11 days after the explosion.  Using measurements of four \h\ lines and five \he\ lines, we show that the H and He line forming regions are separated by \about 4,000~\kms\ at all phases.  We suggest that this gap is consistent with an optically thick, H-rich shell that encloses the explosion and conceals material beneath it.  In SN~2011dh, expansion makes the H-shell transparent and reveals the slower-moving He-rich region about day 11. 

Using the time intervals from the explosion to the first detection of \he, we estimate the relative masses for the hydrogen rich envelopes surrounding three SN~IIb.  We estimate that the mass of the hydrogen shell surrounding the progenitor of SN~2011dh was about three times more massive than the shell around SN~2008ax but about eight times less massive than the shell that enclosed SN~1993J.  

We present photometry of SN~2011dh from 12 filters and construct a bolometric light curve that has a maximum luminosity of $1.8 \pm 0.2 \times 10^{42}$~erg~$s^{-1}$ on day 22.  The NIR fraction of the total bolometric luminosity in SN~2011dh is  31\% on day 4 and it rises to 49\% on day 34.  The UV contribution to the total flux is 16\% on day 4, declining to 5\% on day 9 and to 1\% on day 29.  

We compare the bolometric light curves for SN~IIb~2011dh, SN~IIb~2008ax and SN~IIb~1993J.  The LC display significant differences in the first 12 days after the explosion due to different cooling times of the shocked regions.  At subsequent phases however, the LC for the three SN~IIb are very similar.  Thus the hydrogen shell mass appears to only influence the very early light curves of these SN~IIb.  The agreement of the bolometric luminosities, the times of the maxima, the decline rates and the $B-V$ colors for these three SN~IIb suggests that the progenitors may have had similar masses and composition when they exploded inside hydrogen shells of differing masses and distributions.

The progenitor of SN~2011dh has recently been identified as a yellow supergiant star.  Successful models based on that progenitor will have to match the detailed velocity measurements and light curve data presented here.

\acknowledgments

The CfA Supernova Program is supported by NSF grant AST-1211196 to the Harvard College Observatory.  RPK was supported in part by the National Science Foundation under Grant NSF PHY-1125915 to the Kavli Institute for Theoretical Physics.  J.V. is supported by Hungarian OTKA Grants K-76816 and NN-107637, NSF grant AST-0707769, and Texas Advanced Research Project ARP-009.  JCW is supported by NSF grant AST-1109801.  ASF acknowledges support from a NSF Graduate Research Fellowship and a NASA Graduate Research Program Fellowship.   Support for program GO-12540 was provided by NASA through a grant from the Space Telescope Science Institute, which is operated by the Association of Universities for Research in Astronomy, Inc., under NASA contract NAS5-26555.   KT has also received support from Hungarian OTKA Grant K-76816.  

GHM thanks John Rayner, Alan Tokunaga, Bill Golish, David Griep, Paul Sears, and Eric  Volquardsen at the IRTF for supporting target-of-opportunity observations.  GHM is a visiting Astronomer at the Infrared Telescope Facility, which is operated by the University of Hawaii under Cooperative Agreement no. NNX-08AE38A with the National Aeronautics and Space Administration, Science Mission Directorate, Planetary Astronomy Program.  We thank Gaspar Bakos, David Latham, and Matthew Holman for making FLWO observing time available.  We acknowledge the work of C. Klein on the PAIRITEL mosaic data reduction pipeline and we reference observations made with the Vatican Advanced Technology Telescope.    The authors make frequent use of David Bishop's excellent webpage listing recent supernovae and valuable references associated with them: www.rochesterastronomy.org/snimages/. 

{\it Facilities:}  \facility{HST (STIS)}, \facility{FLWO:1.5m (FAST)}, \facility{HET (LRS)}, \facility{Swift (UVOT; UV grism)}, \facility{IRTF (SpeX)}, \facility{FLWO:1.2m (KepCam)}, \facility{FLWO:PAIRITEL}



\clearpage

\begin{deluxetable}{rccrr}
\tabletypesize{\scriptsize}
\tablecolumns{5} 
\tablewidth{0pc}
\tablecaption{Spectroscopic Observations of SN~2011dh\label{spectbl}}
\tablehead{ \colhead{Date (UT)\tablenotemark{a}} & \colhead{Instrument} & \colhead{Range (\AA)} & \colhead{R} & \colhead{Exp. (s)}}
\startdata
Jun   4.2 & FAST & 3480 - 7420 & 1350 &    900 \\ 
Jun   5.2 & FAST & 3480 - 7420 & 1350 &   1500 \\ 
Jun 6.1  & HET  & 4200 - 10100 &  280 & 900 \\
Jun   6.2 & FAST & 3480 - 7420 & 1350 &   1200 \\ 
Jun   7.2 & FAST & 3480 - 7420 &  1350 &   1200 \\ 
Jun   8.2 & FAST & 3480 - 7420 & 1350 &    900 \\ 
Jun   8.4    &  IRTF  & 8,000 - 24,000  & 1200  &  1800  \\
Jun   9.2 & FAST & 3480 - 7420 & 1350 &    900 \\ 
Jun   9.9  & SWIFT  & 2,000 - 4600 & 150  & 900  \\
Jun 11.1  & SWIFT  & 2,000 - 4600 &  150  & 1135  \\
Jun 11.2 & HET  & 4200 - 10100 &  280 & 900 \\
Jun 12.4   &  IRTF  & 8,000 - 24,000  & 1200  &  1800  \\
Jun 14.2  & HET  & 4200 - 10000 &  280 & 900 \\
Jun 16.4   &  IRTF  & 8,000 - 24,000  & 1200  &  1800  \\
Jun 17.2  & HET  & 4200 - 10100 &  280 & 900 \\
Jun 22.2  & HET  & 4200 - 10100 &  280 & 900 \\
Jun 24.0  & HST & 2160 - 10230 & 4300 & 3600 \\
Jun  25.2 & FAST & 3480 - 7420 & 1350 &    720 \\ 
Jun  26.2 & FAST & 3480 - 7420 & 1350 &    720 \\ 
Jun  27.2 & FAST & 3480 - 7420 & 1350 &    720 \\ 
Jun  28.3 & FAST & 3480 - 7420 & 1350 &    840 \\ 
Jun 29.2  & HET  & 4200 - 10100 &  280 & 900 \\
Jun  30.3 & FAST & 3480 - 7420 & 1350 &    900 \\ 
Jul  1.2 & FAST & 3480 - 7420 & 1350 &    900 \\ 
Jul  2.2 & FAST & 3480 - 7420 & 1350 &    960 \\ 
Jul  4.2 & FAST & 3480 - 7420 & 1350 &    900 \\ 
\enddata
\tablecomments{The time of explosion is \about May 31.5 UT and the time of \emph{B}-max is \about June 20 UT.}
\tablenotetext{a}{All observations in 2011.}
\end{deluxetable}

\begin{deluxetable}{ccccc}
\tabletypesize{\scriptsize}
\tablecolumns{5} 
\tablewidth{0pc}
\tablecaption{Peak Magnitudes and Filter Details \label{peaktbl}}
\tablehead{ \colhead{Band} & \colhead{$M_{peak}$ (mag)} & \colhead{$D_{peak}$\tablenotemark{a}} & \colhead{$N_{obs}$} &
                   \colhead{$\lambda_{mid}$(\AA)} }
\startdata
\cutinhead{PAIRITEL\tablenotemark{b}}
 K$_s$ & $< 11.91\tablenotemark{c} $ & $> 24\tablenotemark{c} $ & 18 & 21590   \\
 H  &       $< 11.91\tablenotemark{c} $ & $> 24\tablenotemark{c} $ & 18 & 16620   \\
 J   &       $< 12.02\tablenotemark{c} $ & $> 24\tablenotemark{c} $ & 17 & 12350  \\
\cutinhead{KeplerCam}
 i'     &     12.45  &  23.2 &  27  & 7747    \\
 r'    &     12.26 &  21.7 &   27 & 6217    \\
 V    &    12.44 &  20.6 &   26 & 5375    \\
 B    &    13.17 &  20.0 &   25 & 4270    \\
 u'   &    14.42 &  16.6 &   14 & 3525    \\
\cutinhead{\emph{Swift}}
 V &  12.63 &  20.9 &       37 & 5468   \\
 B &  13.39 &  19.5 &       39 & 4392    \\
 U &  13.90 &  16.5 &       40 & 3465    \\
 UVW1 &  15.53 &  15.1 &  27 & 2600    \\
 UVM2 &  \nodata &  \nodata &  27 & 2246    \\
 UVW2 &  \nodata &  \nodata &  27 & 1928    \\
\enddata
\tablecomments{Systematic offsets of \emph{u'}$= -0.05$ mag,  \emph{B}$= +0.15$ mag, \emph{V}$= +0.11$ mag, \emph{r'}$= -0.04$ mag and \emph{i'}$= +0.08$ mag are found between the FLWO 1.2m data and other published data (See \S~\ref{photo}.).  Accounting for these offsets,  the FLWO light curves match \citet{Arcavi11} (\emph{i'}) and \citet{Tsvetkov12} (\emph{u'BVr'}) within $\pm 0.05$ mag.}
\tablenotetext{a}{Whole days with respect to the time of explosion.}
\tablenotetext{b}{Filter details from \citet{Cohen03}.}
\tablenotetext{c}{Final measurements on day 24 precede the peak by a day or two.}
\end{deluxetable}

\begin{deluxetable}{ccccccccccc}
\tabletypesize{\scriptsize}
\tablecolumns{11} 
\tablewidth{0pc}
\tablecaption{Photometric Measurements with the FLWO 1.2m and KepCam \label{keptbl}}
\tablehead{ \colhead{Date (UT)\tablenotemark{a}} &
\colhead{u'} & \colhead{u' err } &
\colhead{B} & \colhead{B err } &
\colhead{V} &  \colhead{V err } &
\colhead{r'} & \colhead{r' err } &
\colhead{i'} & \colhead{i' err }}
\startdata
55716.2 & 15.676 &  0.012 & 15.115 &  0.011 & 14.692 &  0.010 & 14.534 &  0.015 & 14.759 &  0.020  \\
55717.2 & 15.754 &  0.023 & 14.866 &  0.009 & 14.363 &  0.008 & 14.186 &  0.010 & 14.510 &  0.013  \\
55718.2 & 15.563 &  0.016 & 14.580 &  0.009 & 14.056 &  0.009 & 13.924 &  0.012 & 14.242 &  0.017  \\
55719.2 & 15.307 &  0.011 & 14.264 &  0.007 & 13.743 &  0.008 & 13.602 &  0.011 & 13.941 &  0.014  \\
55721.1 & 14.971 &  0.154 & 13.849 &  0.006 & 13.305 &  0.010 & 13.165 &  0.016 & 13.543 &  0.020  \\
55722.1 & 14.854 &  0.134 & 13.677 &  0.007 & 13.128 &  0.010 & 13.007 &  0.013 & 13.358 &  0.017  \\
55723.2 & \nodata & \nodata & 13.549 &  0.011 & 13.018 &  0.011 & 12.899 &  0.013 & 13.208 &  0.013  \\
55724.2 & \nodata & \nodata & 13.462 &  0.006 & 12.879 &  0.009 & 12.758 &  0.012 & 13.087 &  0.019  \\
55725.2 & \nodata & \nodata & 13.397 &  0.006 & 12.788 &  0.009 & 12.650 &  0.012 & 12.981 &  0.015  \\
55726.2 & \nodata & \nodata & 13.351 &  0.004 & 12.735 &  0.006 & 12.605 &  0.008 & 12.890 &  0.010  \\
55727.2 & \nodata & \nodata & 13.295 &  0.006 & 12.667 &  0.008 & 12.525 &  0.013 & 12.814 &  0.017  \\
55729.2 & \nodata & \nodata & 13.199 &  0.007 & 12.555 &  0.009 & 12.419 &  0.011 & 12.686 &  0.016  \\
55730.2 & \nodata & \nodata & \nodata & \nodata & 12.514 &  0.008 & 12.391 &  0.010 & 12.634 &  0.012  \\
55731.3 & \nodata & \nodata & 13.165 &  0.006 & 12.490 &  0.008 & 12.342 &  0.010 & 12.587 &  0.014  \\
55732.1 & 14.456 &  0.008 & 13.164 &  0.009 & 12.463 &  0.008 & 12.311 &  0.011 & 12.537 &  0.015  \\
55733.3 & \nodata & \nodata & 13.183 &  0.008 & 12.451 &  0.010 & 12.281 &  0.013 & 12.506 &  0.014  \\
55734.2 & 14.578 &  0.006 & 13.211 &  0.004 & 12.456 &  0.005 & 12.282 &  0.006 & 12.481 &  0.007  \\
55735.2 & \nodata & \nodata & 13.293 &  0.004 & 12.488 &  0.003 & 12.290 &  0.005 & 12.467 &  0.005  \\
55736.2 & 14.877 &  0.010 & 13.386 &  0.008 & 12.515 &  0.010 & 12.315 &  0.011 & 12.495 &  0.014  \\
55738.2 & 15.261 &  0.011 & 13.650 &  0.006 & 12.676 &  0.005 & 12.398 &  0.005 & 12.523 &  0.006  \\
55739.2 & 15.510 &  0.011 & 13.810 &  0.005 & 12.773 &  0.004 & 12.482 &  0.007 & 12.571 &  0.007  \\
55740.2 & \nodata & \nodata & 13.900 &  0.006 & 12.868 &  0.003 & 12.551 &  0.003 & 12.618 &  0.003  \\
55742.2 & 16.059 &  0.018 & 14.236 &  0.009 & 13.069 &  0.010 & 12.687 &  0.011 & 12.727 &  0.016  \\
55743.2 & 16.215 &  0.024 & 14.324 &  0.008 & 13.169 &  0.004 & 12.765 &  0.003 & 12.756 &  0.003  \\
55744.2 & 16.443 &  0.026 & 14.474 &  0.008 & 13.249 &  0.006 & 12.829 &  0.008 & 12.818 &  0.010  \\
\enddata
\tablecomments{Systematic offsets of \emph{u'}$= -0.05$ mag,  \emph{B}$= +0.15$ mag, \emph{V}$= +0.11$ mag, \emph{r'}$= -0.04$ mag and \emph{i'}$= +0.08$ mag are found between the FLWO 1.2m data and other published data (See \S~\ref{photo}.).  Accounting for these offsets,  the FLWO light curves match \citet{Arcavi11} (\emph{i'}) and \citet{Tsvetkov12} (\emph{u'BVr'}) within $\pm 0.05$ mag.}
\tablenotetext{a}{Average time of observation for a sequence with multiple filters.  Estimated date of explosion: MJD=55712.5}
\end{deluxetable}

\begin{deluxetable}{ccccccc}
\tabletypesize{\scriptsize}
\tablecolumns{7} 
\tablewidth{0pc}
\tablecaption{Photometric Measurements with PAIRITEL\label{pteltbl}}
\tablehead{ \colhead{Date (UT)\tablenotemark{a}} & 
\colhead{J} & \colhead{J err } &
\colhead{H} & \colhead{H err } &
\colhead{$K_{s}$} & \colhead{$K_{s}$ err }}
\startdata
55717.1 & 13.875 &  0.193 & 13.762 &  0.535 & 13.607 &  0.369  \\
55718.2 & 13.624 &  0.153 & 13.473 &  0.401 & 13.497 &  0.341  \\
55719.2 & 13.398 &  0.123 & 13.400 &  0.370 & 13.077 &  0.255  \\
55720.2 & 13.196 &  0.100 & 13.154 &  0.292 & 13.027 &  0.296  \\
55721.2 & 12.993 &  0.082 & 13.017 &  0.255 & 12.969 &  0.185  \\
55722.2 & 12.821 &  0.068 & 12.807 &  0.205 & 12.702 &  0.151  \\
55723.2 & 12.696 &  0.061 & 12.659 &  0.178 & 12.510 &  0.153  \\
55724.2 & 12.585 &  0.055 & 12.532 &  0.155 & 12.317 &  0.122  \\
55725.1 & 12.477 &  0.050 & 12.473 &  0.151 & 12.427 &  0.125  \\
55726.2 & 12.377 &  0.054 & 12.375 &  0.148 & \nodata &  \nodata  \\
55727.2 & 12.327 &  0.050 & 12.291 &  0.130 & 12.214 &  0.162  \\
55728.2 & 12.230 &  0.044 & 12.140 &  0.119 & 12.047 &  0.156  \\
55730.2 & 12.143 &  0.041 & 12.102 &  0.110 & 11.933 &  0.133  \\
55731.2 & 12.105 &  0.038 & 12.016 &  0.097 & 12.,000 &  0.101  \\
55732.2 & 12.039 &  0.036 & 12.045 &  0.105 & 11.934 &  0.108  \\
55733.2 & 12.051 &  0.036 & 11.908 &  0.088 & 11.986 &  0.030  \\
55734.2 & 11.868 &  0.081 & 11.868 &  0.077 & 11.976 &  0.039  \\
55735.2 & 11.898 &  0.099 & 11.931 &  0.217 & 11.931 &  0.217  \\
\enddata
\tablenotetext{a}{Average time of observation for a sequence with multiple filters.  Estimated date of explosion: MJD=55712.5}
\end{deluxetable}

\begin{deluxetable}{lcccccccccccc}
\tabletypesize{\scriptsize}
\tablecolumns{13} 
\tablewidth{0pc}
\tablecaption{Sequence of Local Comparison stars \label{localcomp}}
\tablehead{\colhead{ID} & \colhead{R.A.(J2000)} & \colhead{Dec.(J2000)} & 
\colhead{\emph{u'}} & \colhead{\emph{u'}} err & \colhead{\emph{B}} & \colhead{\emph{B}} err & 
\colhead{\emph{V}} & \colhead{\emph{V}} err & \colhead{\emph{r'}} & \colhead{\emph{r'}} err & \colhead{\emph{i'}} & \colhead{\emph{i'} err}}
\startdata
A & 13:30:21.93 & +47:09:05.7 & 17.230 & 0.098 & 16.454 & 0.021 & 15.895 & 0.014 & 15.686 & 0.017 & 15.495 & 0.023 \\
B & 13:30:14.29 & +47:14:56.0 & 17.922 & 0.110 & 16.918 & 0.021 & 16.295 & 0.012 & 16.052 & 0.016 & 15.839 & 0.018 \\
C & 13:30:13.86 & +47:07:30.1 & 17.877 & 0.092 & 16.475 & 0.020 & 15.653 & 0.012 & 15.334 & 0.016 & 15.064 & 0.016 \\
\enddata
\tablecomments{Finding chart is displayed as Figure~\ref{fc}.}
\end{deluxetable}

\begin{deluxetable}{cllllllll}
\tabletypesize{\scriptsize}
\tablecolumns{9} 
\tablewidth{0pc}
\tablecaption{Photometric Measurements with \emph{Swift}\label{swifttbl}}
\tablehead{ \colhead{Date (UT)\tablenotemark{a}} & 
\colhead{UVW2} & \colhead{W2 err} &
\colhead{UVM2} & \colhead{M2 err} & 
\colhead{UVW1} & \colhead{W1 err} &  
\colhead{U} & \colhead{U err}}
\startdata
55716.0 & 16.31 & 0.03 & 15.94 & 0.03 & 15.40 & 0.02 & 14.95 & 0.02 \\
55716.7 & 16.61 & 0.05 & 16.19 & 0.05 & 15.56 & 0.03 & 15.09 & 0.02  \\
55717.7 & 16.83 & 0.07 & 16.51 & 0.08 & 15.81 & 0.05 & 15.14 & 0.03  \\
55719.0 & 16.74 & 0.06 & 16.70 & 0.04 & 15.86 & 0.04 & 14.88 & 0.02  \\
55720.6 & 17.00 & 0.14 & 16.84 & 0.17 & 15.76 & 0.04 & 14.47 & 0.02  \\
55721.8 & 16.80 & 0.08 & 17.19 & 0.11 & 15.68 & 0.03 & 14.28 & 0.02  \\
55723.1 & 16.93 & 0.13 & 17.23 & 0.13 & 15.59 & 0.03 & 14.14 & 0.02  \\
55723.8 & 16.78 & 0.14 & 17.38 & 0.17 & 15.67 & 0.05 & 14.07 & 0.02  \\
55725.1 & 16.89 & 0.05 & 17.16 & 0.08 & 15.63 & 0.03 & 13.99 & 0.02  \\
55726.6 & 16.98 & 0.05 & 17.60 & 0.12 & 15.59 & 0.03 & 14.01 & 0.02  \\
55728.8 & 16.97 & 0.10 & 17.66 & 0.16 & 15.60 & 0.06 & 13.93 & 0.05  \\
55729.2 & 16.86 & 0.10 & 17.42 & 0.14 & 15.61 & 0.06 & 13.87 & 0.05  \\
55730.3 & 16.80 & 0.08 & 17.41 & 0.11 & 15.56 & 0.06 & 13.89 & 0.05  \\
55731.5 & 16.89 & 0.10 & 17.62 & 0.16 & 15.53 & 0.06 & 13.91 & 0.05  \\
55732.5 & 16.89 & 0.08 & 17.66 & 0.12 & 15.63 & 0.06 & 13.93 & 0.05  \\
55733.5 & 17.01 & 0.09 & 17.67 & 0.13 & 15.64 & 0.06 & 13.99 & 0.05  \\
55734.4 & 17.03 & 0.09 & 17.59 & 0.13 & 15.70 & 0.06 & 14.08 & 0.05  \\
55735.7 & 16.97 & 0.09 & 17.81 & 0.15 & 15.82 & 0.07 & 14.25 & 0.05  \\
55736.2 & 17.17 & 0.10 & 17.96 & 0.18 & 15.84 & 0.07 & 14.31 & 0.05  \\
55737.2 & 17.45 & 0.12 & 17.80 & 0.20 & 16.03 & 0.08 & 14.56 & 0.05  \\
55738.8 & 17.39 & 0.09 & 17.79 & 0.31 & 16.24 & 0.05 & 14.82 & 0.03  \\
55740.2 & 17.70 & 0.13 & 18.77 & 0.58 & 16.51 & 0.05 & 15.13 & 0.03  \\
55741.4 & 17.77 & 0.15 & 18.29 & 0.26 & 16.67 & 0.08 & 15.34 & 0.04  \\
55741.8 & 17.89 & 0.15 & 18.43 & 0.25 & 16.66 & 0.07 & 15.49 & 0.04  \\
55743.8 & 18.47 & 0.26 & 19.15 & 0.51 & 17.10 & 0.11 & 15.90 & 0.05  \\
55745.2 & 18.11 & 0.19 & 18.95 & 0.41 & 17.02 & 0.10 & 16.02 & 0.06  \\
55746.1 & 18.54 & 0.27 & 18.32 & 0.22 & 17.31 & 0.12 & 16.18 & 0.06  \\
\enddata
\tablenotetext{a}{Average time of observation for a sequence with multiple filters.  Estimated date of explosion: MJD=55712.5}
\end{deluxetable}

\begin{deluxetable}{rlllllllll}
\tabletypesize{\scriptsize}
\tablecolumns{10} 
\tablewidth{0pc}
\tablecaption{Velocity Measurements of H and He Lines (\kms) \label{vtbl1}}
\tablehead{ \colhead{Epoch\tablenotemark{a}} & \colhead{\ha} & \colhead{H$\beta$} & \colhead{H$\gamma$} & \colhead{Pa$\beta$} &  \colhead{\he\ 5876} & \colhead{\he\ 6678} & \colhead{\he\ 7065} & \colhead{\he\ 1.0830} & \colhead{\he\ 2.0581}}
\startdata
4	&	15400	&	13800	&	14100	&	\nodata	&	\nodata	&	\nodata	&	\nodata	&	\nodata	&	\nodata	 \\
5	&	14500	&	13,000	&	13900	&	\nodata	&	\nodata	&	\nodata	&	\nodata	&	\nodata	&	\nodata	 \\
6	&	14100	&	12700	&	13600	&	\nodata	&	\nodata	&	\nodata	&	\nodata	&	\nodata	&	\nodata	 \\
7	&	13700	&	12200	&	13500	&	\nodata	&	\nodata	&	\nodata	&	\nodata	&	\nodata	&	\nodata	 \\
8	&	13500	&	12,000	&	13400	&	\nodata	&	\nodata	&	\nodata	&	\nodata	&	\nodata	&	\nodata	 \\
9	&	13300	&	11600	&	13100	&	\nodata	&	\nodata	&	\nodata	&	\nodata	&	\nodata	&	\nodata	 \\
10	&	\nodata	&	\nodata	&	12600	&	\nodata	&	\nodata	&	\nodata	&	\nodata	&	\nodata	&	\nodata	 \\
11	&	13100	&	11,000	&	12500	&	\nodata	&	8400	 &	\nodata	&	\nodata	&	\nodata	&	\nodata	 \\
12	&	\nodata	&	\nodata	&	\nodata	&	12200	&	\nodata	&	\nodata	&	\nodata	&	8200	 &    	\nodata	 \\
13	&	\nodata	&	\nodata	&	\nodata	&	\nodata	&	\nodata	&	\nodata	&	\nodata	&	\nodata	&	\nodata	 \\
14	&	12500	&	10800	&	\nodata	&	\nodata	&	7600	&	\nodata	&	\nodata	&	\nodata	&	\nodata	 \\
15	&	\nodata	&	\nodata	&	\nodata	&	12,000	&	\nodata	&	\nodata	&	\nodata	&	\nodata	&	\nodata	 \\
16	&	\nodata	&	\nodata	&	\nodata	&	\nodata	&	\nodata	&	\nodata	&	\nodata	&	8,000	 &	7200	 \\
17	&	12400	&	10700	&	\nodata	&	\nodata	&	7500	&	\nodata	&	\nodata	&	\nodata	&	\nodata	 \\
18	&	\nodata	&	\nodata	&	\nodata	&	11900	&	\nodata	&	\nodata	&	\nodata	&	\nodata	&	\nodata	 \\
19	&	\nodata	&	\nodata	&	\nodata	&	\nodata	&	\nodata	&	\nodata	&	\nodata	&	\nodata	&	\nodata	 \\
20	&	\nodata	&	\nodata	&	\nodata	&	\nodata	&	\nodata	&	\nodata	&	\nodata	&	\nodata	&	\nodata	 \\
21	&	\nodata	&	\nodata	&	\nodata	&	\nodata	&	\nodata	&	\nodata	&	\nodata	&	\nodata	&	\nodata	 \\
22	&	12100	&	10600	&	\nodata	&	\nodata	&	7500	&	6500	&	\nodata	&	\nodata	&	\nodata	 \\
23	&	\nodata	&	\nodata	&	\nodata	&	\nodata	&	\nodata	&	\nodata	&	\nodata	&	\nodata	&	\nodata	 \\
24	&	12100	&	10900	&	\nodata	&	\nodata	&	7600	&	6600	&	7400	&	\nodata	&	\nodata	 \\
25	&	11900	&	11,000	&	\nodata	&	\nodata	&	7400	&	6600	&	7200	&	\nodata	&	\nodata	 \\
26	&	12100	&	11,000	&	\nodata	&	\nodata	&	7600	&	6900	&	7200	&	\nodata	&	\nodata	 \\
27	&	12,000	&	10800	&	\nodata	&	\nodata	&	7600	&	7100	&	7100	&	\nodata	&	\nodata	 \\
28	&	11900	&	10800	&	\nodata	&	\nodata	&	7500	&	7200	&	7100	&	\nodata	&	\nodata	 \\
29	&	12,000	&	10700	&	\nodata	&	\nodata	&	7400	&	7600	&	7,000	&	\nodata	&	\nodata	 \\
30	&	12,000	&	10900	&	\nodata	&	\nodata	&	7500	&	7300	&	6900	&	\nodata	&	\nodata	 \\
31	&	12,000	&	10900	&	\nodata	&	\nodata	&	7600	&	7300	&	7,000	&	\nodata	&	\nodata	 \\
32	&	11900	&	10800	&	\nodata	&	\nodata	&	7700	&	7200	&	7,000	&	\nodata	&	\nodata	 \\
33	&	\nodata	&	\nodata	&	\nodata	&	\nodata	&	\nodata	&	\nodata	&	\nodata	&	\nodata	&	\nodata	 \\
34	&	11800	&	10800	&	\nodata	&	\nodata	&	7800	&	7100	&	6800	&	\nodata	&	\nodata	 \\
\enddata
\tablenotetext{a}{Whole days with respect to the time of explosion.}
\end{deluxetable}

\begin{deluxetable}{rllllll}
\tabletypesize{\scriptsize}
\tablecolumns{7} 
\tablewidth{0pc}
\tablecaption{Velocity Measurements of \ca\ and \fe\ Lines (\kms) \label{vtbl2}}
\tablehead{ \colhead{Epoch\tablenotemark{a}} & \colhead{\ca\ 3945} & \colhead{\ca\ 8579} & \colhead{\fe\ 4561} & \colhead{\fe\ 5018} & \colhead{\fe\ 5169} & \colhead{\fe\ 1500}}
\startdata
4	&	13700	&	\nodata	&	12700	&	\nodata	& 11400  &   \nodata  \\
5	&	12500	&	\nodata	&	12300	&	12200	& 11,000  &   \nodata  \\
6	&	12600	&	14300	&	11800	&	11500	& 10900  &   \nodata  \\
7	&	12500	&	\nodata	&	11700	&	11200	&  10700  &  \nodata  \\
8	&	12400	&	\nodata	&	11900	&	10900	&  10100  &   9500  \\
9	&	12100	&	\nodata	&	11400	&	10700	&  10000  &   \nodata  \\
10	&	11800	&	\nodata	&	\nodata	&	\nodata	&  \nodata  &  \nodata  \\
11	&	11800	&	13,000	&	10500	        &	8800  	&   8800  &  \nodata  \\
12	&	\nodata	&	\nodata	&	\nodata	&	\nodata	&  \nodata  &  9100  \\
13	&	\nodata	&	\nodata	&	\nodata	&	\nodata	&  \nodata  &  \nodata  \\
14	&	\nodata	&	12700	&	8900	        &	8,000  	&  7800  &  \nodata  \\
15	&	\nodata	&	\nodata	&	\nodata	&	\nodata	& \nodata  &   \nodata  \\
16	&	\nodata	&	\nodata	&	\nodata	&	\nodata	&  \nodata  &  8700  \\
17	&	\nodata	&	12100	&	8800	        &	7600  	&   7500   &  \nodata  \\
18	&	\nodata	&	\nodata	&	\nodata	&	\nodata	&  \nodata  &  \nodata  \\
19	&	\nodata	&	\nodata	&	\nodata	&	\nodata	&  \nodata   &  \nodata \\
20	&	\nodata	&	\nodata	&	\nodata	&	\nodata	&  \nodata   &  \nodata \\
21	&	\nodata	&	\nodata	&	\nodata	&	\nodata	&  \nodata   &  \nodata \\
22	&	\nodata	&	10800	&	8200  	&	6900	         &    6900    & \nodata  \\
23	&	\nodata	&	\nodata	&	\nodata	&	\nodata	&  \nodata  &  \nodata  \\
24	&     10600	       &   10000  	&	7500 	&	6800	        &   6900   &  \nodata  \\
25	&     10800	      &	\nodata	&	7300	        &	6600 	 &  6600   &  \nodata  \\
26	&     10700	      &	\nodata	&	6900	        &	6500 	 &  6400   &  \nodata  \\
27	&	10700	      &	\nodata	&	\nodata	&	\nodata     &  6400   &   \nodata  \\
28	&	10300	      &	\nodata	&	\nodata    &	\nodata     &   6100   &  \nodata  \\
29	&	\nodata	      &    9600	 &	\nodata        &	\nodata     &   \nodata   &  \nodata  \\
30	&	9900	      &	\nodata	&	\nodata        &	\nodata	&   6,000  &   \nodata  \\
31	&	9800	      &	\nodata	&	\nodata    &	\nodata	&   5800  &  \nodata  \\
32	&	9500	      &	\nodata	&	\nodata    &	\nodata	&  5700   &  \nodata  \\
33	&	\nodata   &	\nodata	&	\nodata	&	\nodata	&  \nodata  &  \nodata  \\
34	&	9200	      &	\nodata	&	\nodata    &	\nodata	&  5300  &  \nodata  \\
\enddata
\tablenotetext{a}{Whole days with respect to the time of explosion.}
\end{deluxetable}

\begin{deluxetable}{rccccc}
\tabletypesize{\scriptsize}
\tablecolumns{6} 
\tablewidth{0pc}
\tablecaption{Bolometric Luminosity by Phase \label{boltbl}}
\tablehead{ \colhead{Epoch\tablenotemark{a}} & \colhead{Total Flux \tablenotemark{b}} & 
                  \colhead{NIR Flux} & \colhead{NIR} & \colhead{UV Flux} & \colhead{UV} \\
\colhead{(Days)} & \colhead{($10^{42}$ erg s$^{-1}$)} & 
                  \colhead{($10^{42}$ erg s$^{-1}$)} & \colhead{Fraction} & \colhead{($10^{42}$ erg s$^{-1}$)} & \colhead{Fraction}}
\startdata
  4 &   0.29 &   0.09 &   0.31 &   0.05 &   0.16\\
  5 &   0.35 &   0.11 &   0.31 &   0.04 &   0.11\\
  6 &   0.43 &   0.14 &   0.32 &   0.03 &   0.08\\
  7 &   0.56 &   0.17 &   0.31 &   0.04 &   0.06\\
  8 &   0.67 &   0.21 &   0.31 &   0.04 &   0.06\\
  9 &   0.80 &   0.24 &   0.31 &   0.04 &   0.05\\
 10 &   0.94 &   0.30 &   0.32 &   0.05 &   0.05\\
 11 &   1.05 &   0.33 &   0.32 &   0.05 &   0.05\\
 12 &   1.17 &   0.37 &   0.32 &   0.05 &   0.04\\
 13 &   1.27 &   0.41 &   0.32 &   0.05 &   0.04\\
 14 &   1.34 &   0.44 &   0.33 &   0.05 &   0.04\\
 15 &   1.43 &   0.47 &   0.33 &   0.05 &   0.04\\
 16 &   1.52 &   0.51 &   0.34 &   0.06 &   0.04\\
 17 &   1.59 &   0.54 &   0.34 &   0.06 &   0.04\\
 18 &   1.65 &   0.56 &   0.34 &   0.06 &   0.04\\
 19 &   1.69 &   0.58 &   0.34 &   0.06 &   0.03\\
 20 &   1.74 &   0.61 &   0.35 &   0.06 &   0.03\\
 21 &   1.77 &   0.62 &   0.35 &   0.05 &   0.03\\
 22 &   1.78 &   0.64 &   0.36 &   0.05 &   0.03\\
 23 &   1.78 &   0.68 &   0.38 &   0.04 &   0.03\\
 24 &   1.73 &   0.67 &   0.39 &   0.04 &   0.02\\
 25 &   1.65 &   0.67 &   0.40 &   0.03 &   0.02\\
 26 &   1.58 &   0.66 &   0.42 &   0.03 &   0.02\\
 27 &   1.48 &   0.63 &   0.43 &   0.03 &   0.02\\
 28 &   1.39 &   0.61 &   0.44 &   0.02 &   0.02\\
 29 &   1.29 &   0.58 &   0.45 &   0.02 &   0.01\\
 30 &   1.21 &   0.55 &   0.46 &   0.02 &   0.01\\
 31 &   1.14 &   0.54 &   0.47 &   0.01 &   0.01\\
 32 &   1.07 &   0.51 &   0.47 &   0.01 &   0.01\\
 33 &   1.01 &   0.49 &   0.48 &   0.01 &   0.01\\
 34 &   0.97 &   0.47 &   0.49 &   0.01 &   0.01\\
 \enddata
\tablenotetext{a}{Whole days with respect to the time of explosion.}
\tablenotetext{b}{Total measurement uncertainties are \about 10\% of the bolometric flux.}
\end{deluxetable}

\end{document}